# Multivariate Functional Data Analysis Uncovers Behavioural Fingerprints in Invertebrate Locomotor Response to Micropollutants

**Authors:**
George Ruck[1,2], Jean-Baptiste Aubin[3], Alexandre Delétang[2], Didier Neuzeret[2], Olivier Geffard[1], Arnaud Chaumot[1]*

[1]INRAE, UR RiverLy, Laboratoire d'écotoxicologie, F-69625 Villeurbanne, France
[2]Viewpoint, 67 rue Copernic, F-01390 Civrieux, France
[3]University of Lyon, INSA Lyon, Laboratory DEEP – EA 7429, 11 rue de la physique, F-69621 Villeurbanne
*corresponding author: arnaud.chaumot@inrae.fr

**Synopsis**
Functional Data Analysis and machine learning interpret locomotor responses of three aquatic invertebrates, revealing multispecies behavioural fingerprints for micropollutants in both lab tests and WWTP surge events.

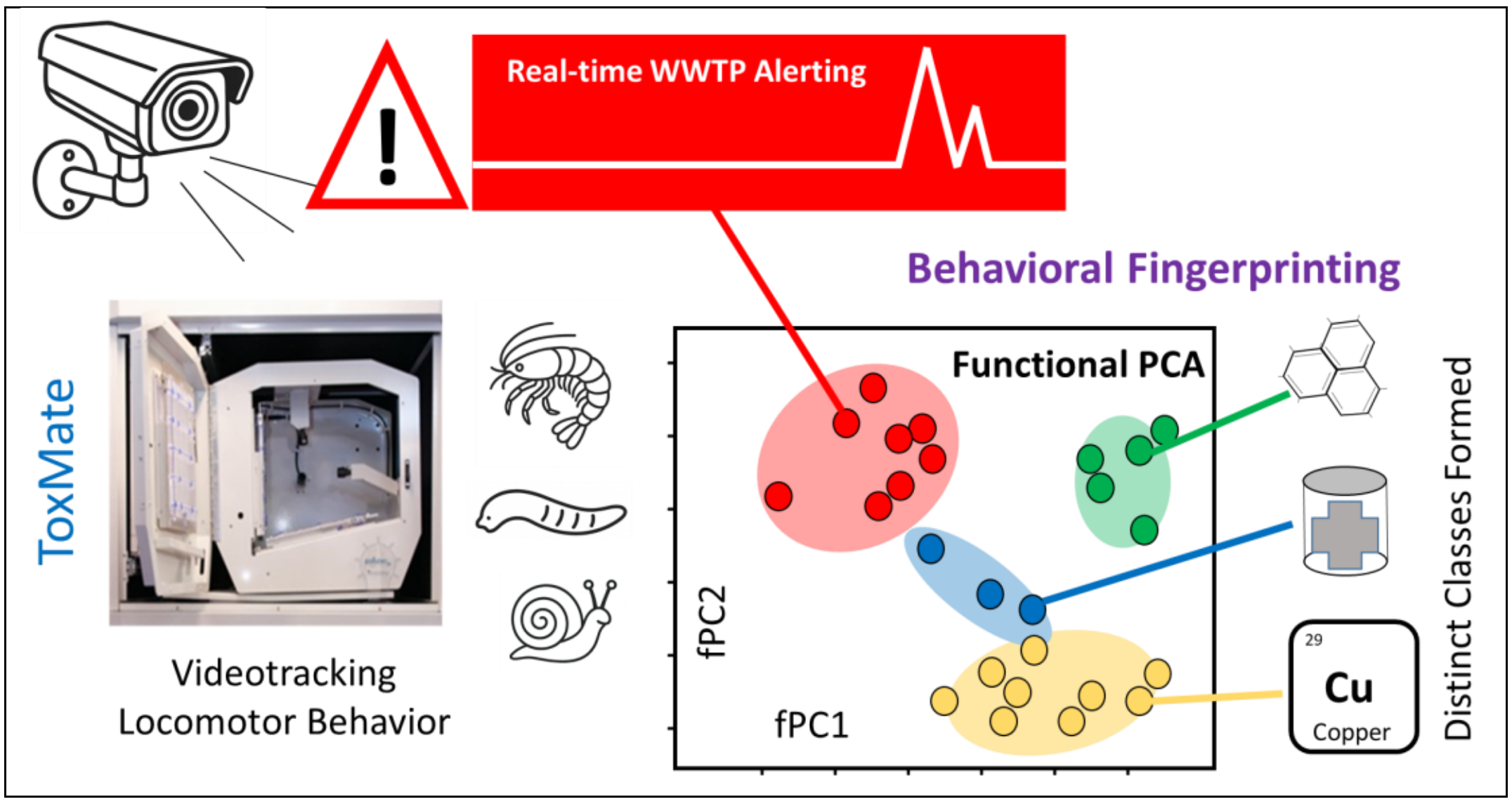

## Abstract

The need for effective biomonitoring in wastewater has become clear due to the impracticality of continuously tracking all chemicals and emerging contaminants in the aquatic exposome. Effect-based biomonitoring provides a cost-effective solution. The ToxMate device, which uses videotracking of locomotor behavior in aquatic invertebrates, has proven efficient for real-time detection of micropollutant surges in effluents. To extend the approach, this proof-of-concept study evaluates the potential to formalize behavioral fingerprints from real-time videotracking data to characterize qualitative variations in effluent contamination. We present the first application of a functional data analysis (FDA) framework in ecotoxicology. Data were obtained by simultaneously tracking three sentinel organisms from distinct taxa (a crustacean, an annelid, and a gastropod) during pulse exposures to four chemicals in the laboratory (two metals, one pharmaceutical, and one insecticide). Individual and multispecies responses were analy-zed to determine whether combining species enhances the resolution of contamination fingerprints through multidimensional FDA. Applying the same data-driven approach to field data from a wastewater treatment plant (WWTP) revealed four recurring types of micropollution events. This proof of concept demonstrates the potential of behavioral fingerprints to improve wastewater monitoring and reduce pollutant transfer to the environment.

## 1. Introduction

Wastewater treatment plants (WWTPs) are the converging point of the majority of urban and industrial micropollutants, designed for the elimination of harmful contaminants prior to discharge to the aquatic environment.[1] Despite efforts to remove micropollutants, concentrations vary greatly at both WWTP inlets and outlets, and micropollutant surge is repeatedly observed through sampling campaigns and across European studies.[2,3] Pulse like episodes of micropollutant surge, which pose known threats to biodiversity and human health,[4,5] can occur over timeframes as short as a couple of hours, making them particularly difficult to identify.[6-8] The passage of such events may either go unnoticed or show up through latent effects,[9] once already present in the aquatic environment. To address this problem, behavioural biomonitoring has been shown as an effective real-time alerting system at WWTP outlets, making use of sublethal organism sensitivity and quick response times to toxic compounds.[10-12] Whilst such biomarkers have long been identified, recent advances in technologies such as Machine Learning and videotracking software have opened the door to more detailed interpretation.[13-14] Our previous work presented an effective bioassay for micropollutant surge detection and its real-time biomonitoring application,[14] developed to tackle the problems associated with the dynamic nature of micropollutant surge, with the device ToxMate.[15] The ToxMate is currently operational at wastewater monitoring sites in France, and whilst alerting indicates the presence of potential issues, it provides limited information on the chemical nature of the pollution.

In a pioneering behavioural study on high-throughput screening for psychoactive drug discovery, Rihel et al. showed that locomotion responses of the model species *Danio rerio* could be clustered into groups of distinct behavioural phenotypes: behavioural fingerprints.[16] More recent studies with *D. rerio* have also shown behavioural sensitivity to low concentrations of other micropollutants common to WWTPs, such as pharmaceuticals, metals and even PFAS.[17-19] As pointed out in our first study,[14] there were distinct differences between the *Gammarus fossarum* avoidance response to metal and pesticide contamination, which could similarly be used to reveal information on the nature of unknown surge toxicity events at monitoring sites. Behavioural profiling in zebrafish offers a cost-effective way to classifiy neuroactive drugs and predict biological targets of novel compounds, and we hypothesise that the same approach could be a cost effective way for characterising surge contamination at effluents.

By means of advanced data science approaches, the present study looks to develop previous work by assessing the feasibility of such a behavioural profiling of contaminated waters. As the ToxMate device was developed with three parallel tracking panels (16 individual chambers per panel), this study extends our previous single species *G. fossarum* approach, to a multispecies approach, incorporating two other novel species of macroinvertebrate: the leech *Erpobdella testacea* and the aquatic snail *Radix auricularia.* The new taxa were specifically selected for their compatibility with the experimental apparatus and their taxonomic distance from *Gammarus*, which is anticipated to provide contrasting sensitivities to various contaminants. It is expected that adding different phyla to complement the crustacean model broadens detection scope through varying species sensitivities, will make the biotest more suited to diverse effluents that may contain any groups of commonly evaluated compounds such as metals and pesticides as well as contaminants of emerging concern (CECs).[20,21] Furthermore, the presence of multiple species adds signal multidimensionality that is shown to be advantageous for clustering algorithms and toxicity evaluation.[16,22] To be clear, the reference definition in this paper's context of a behavioural fingerprint is therefore the multivariate trajectory of locomotor behavioural patterns over time, across the multiple invertebrate species. Our study's primary objective was to establish a data-driven statistical framework to formalize this behavioural fingerprinting based on videotracking data. First, we provide a proof of concept using single-substance laboratory exposures to develop the statistical framework, define relevant metrics, and introduce a novel multivariate approach adapted to the problem (multidimensional Functional Data Analysis). Second, we apply the developed statistical framework to field data collected on a WWTP effluent (completely independently from the lab-based fingerprint data), demonstrating that this multivariate analysis of multi-species behavioural signals enables forming behavioural fingerprinting of the effluent that captures temporal variability of chemical contamination.

The data science framework chosen to cluster behavioural fingerprints is functional data analysis (FDA),[23,24] never employed in ecotoxicology to our knowledge. FDA deviates from traditional multivariate statistics by treating observations as functions, such as curves or surfaces, where each function portrays a continuous and often smooth process as random variables. This is well suited to time series data, as it preserves the sequential nature of the data in extracted features, typically through a set of functions in a basis domain, such as B-spline polynomials or Fourier functions.[23] Where FDA was once computationally heavy, the calculation of intrinsic mathematical properties such as the functional covariance and Eigen functions has become straightforward with modern computing, making FDA well suited to exploratory techniques like dimension reduction (functional principal component analysis (fPCA)) and unsupervised classification methods such as k-means clustering. The availability of robust packages in R and the emergence of Python packages like scikit-fda, mirroring popular machine learning packages such as scikit-learn, have contributed to the increasing adoption of FDA.[24] Furthermore, recent FDA research has largely extended univariate techniques to multivariate data,[25,26] making it suitable for a multispecies approach.

To demonstrate the observation of distinct behavioural fingerprints between WWTP relevant micropollutants, in addition to the previously tested copper (Cu) and methomyl, the metal zinc (Zn) and the pharmaceutical verapamil were also tested in the laboratory with the spike exposure methodology described in Ruck et al (2023).[14] Zn was chosen for its high industrial use and known concentration surge in influents and effluents.[2,27,28] Verapamil, a pharmaceutical compound previously shown to trigger locomotor behavioural response,[16] represents the pharmaceuticals class of micropollutants. Pharmaceuticals are often identified as the most abundant micropollutants in effluents,[29,30] and known for their persistence in treatment processes,[20] mass industrial production,[31] and toxicity at lower concentrations than previously thought.[32] In the second part of this study, we applied the same functional machine learning methodology on field data from a long-term on-site WWTP effluent monitoring site in north western France. By first characterising distinct behavioural fingerprints in controlled laboratory conditions, we then tested the transferability of this data-driven statistical approach to real-time field monitoring, aiming to identify recurring types of micropollutant surges in urban effluents. As a complementary exploratory step, automatic grab samples were triggered by spikes in the bioactivity signal to examine the chemical coherence of grouped field events.

Without a device such as the ToxMate, sampled concentrations of measured compounds are bound to inferior to true maximum values at effluents, as the sampling is performed without real-time indication of the micropollutant transfer. Each grab sample was analysed for a total of 803 organic micropollutants (semi-quantitative analysis), primarily based on pesticides, pharmaceuticals and their associated transformation products (TPs), known by local water authorities to be prevalent in this area. We explore the possible alignment of similarities in micropollutant mixture chemical profiles with overlapping behavioural fingerprints.

# 2. Methods

## 2.1 Experimental design and setup

All experimental conditions remain unchanged from the procedure outlined in our previous study,[14] and the presented results are extended for all three species of invertebrate. Briefly, the experimental device comprises a compact, cubic monitoring unit equipped with three vertical observation panels, each integrating a 4 × 4 array of individual behavioural chambers, enabling high-throughput, continuous monitoring under dynamic flow conditions (Fig. 1). The 48 chambers are hydraulically connected to a central reservoir, forming a closed-loop recirculating system in laboratory experiments with a total water volume of 15 L. Each individual arena has internal dimensions of 55 × 50 × 18 mm. For online monitoring of WWTP effluents, the system is modified with dedicated inlet and outlet ports allowing continuous bypass flow-through of raw effluent. Additionally, the setup incorporates a temperature control module with a cooling chamber and a pre-filtration unit prior to water distribution within the ToxMate device. The calibration and normalisation applied for *E. testacea* and *R. auricularia*, as well as a reminder of the spike exposure protocol are given in Appendix 0 of the supplementary information. The amphipod *G. fossarum*, the freshwater leech *E. testacea* and the aquatic snail *R. auricularia,* were all sampled in a former watercress farm near Saint-Maurice-de-Rémens in Central Eastern France (45°57'25"N 5°15'42"E). Hence, the gammarids used in this study originate from the reference source population that has been used in all ecotoxicological experiments conducted by the INRAE laboratory since 2015. Although not specifically characterized for this study, this population has been genetically analyzed on multiple occasions. COI barcoding identified the majority of individuals as *Gammarus fossarum* lineage B (MOTU 2), with a minor co-occurrence of individuals from lineage C (MOTU 6), following the nomenclature from Wattier et al (2020).[33] No genetic information is available for the two other species. All species adapted well to the previously described acclimatisation conditions at the laboratory in Lyon, France. Each of the three species is placed in one of the three panels of the ToxMate device, with each panel allowing the monitoring of 16 individuals housed in separate chambers. The four tested molecules were copper (CAS 10125-13-0; 8 replicated experiments), methomyl (CAS 16752-77-5; 7 replicated experiments), zinc (CAS 7446-20-0; 4 replicated experiments), verapamil (CAS 52-53-9; 3 replicated experiments), adopting the same 2h spike exposure experimental design outlined in Ruck et al (2023).[14] Concentrations (100 $\mu gL^{-1}$ for copper, 125 $\mu gL^{-1}$ for methomyl, 325 $\mu gL^{-1}$ for zinc, 120 $\mu gL^{-1}$ for verapamil) were tested to remain sublethal, and when possible approximated to be a fifth of the LC50 for *G. fossarum* (as in [14]), the goal being to trigger behavioural response rather than reproduce realistic field conditions. For the data described relative to operational WWTP monitoring, ToxMate biomonitoring has been in operation at the site since January 2022, which is a municipal WWTP in the north western region of France serving an equivalent population of 170,000. Alert triggered grab sampling was activated at the site for two different month-long periods: November 2022 and April 2023, where organisms are continuously exposed to the undiluted effluent. For all field monitoring periods, organisms are replaced monthly across all species to ensure responsiveness is maintained.

The locomotor data collected for behavioural fingerprinting was of the same nature as the data described in our previous work.[14] For each repetition, the trajectories of *N*=16 individuals of each of the three species were exported from the videotracking software developed by Viewpoint, the ToxMate developers. Experimental conditions and water flow setup remained unchanged from Ruck et al (2023).[14] For all repetitions, an initial acclimation of 72 hours in the ToxMate was carried out, and videotracking took place from 24 hours prior to spike exposure, until 48 hours post exposure. In this study, only the two hours post exposure were retained. Of all the retained observations, the dataset is comprised of a total $M$=22 behavioural response assays, encompassing the individual locomotor activity tracking of 1056 organisms (three species, 22 assays). Cu and methomyl, previously tested in [14], contain twice as many assay repetitions, as these chemicals were used heavily in the initial development of the ToxMate spike assay protocol. Any organisms that showed mortality prior to spike testing were automatically filtered through inactivity monitoring.

## 2.2 Data Pre-processing

For each invertebrate species, time series data is composed of distance measures, travelled over 20 second intervals, $X_i(t_j)$, where $i$ represents each of the $N$=1-16 individuals per species, and $j$ the $j$=[0,20,…, $T$ =7200]-second sampling points in the 2-hour time domain $T$. The cumulative distance measure are automatically calculated from centre of gravity coordinates at 10 frame per second (fps) sampling using in-house software of the company Viewpoint (ToxMate supplier). The mean locomotor response ($\mu(t_j)$) and avoidance response ($\alpha(t_j)$ as in Ruck et al (2023)[14] were then computed for each of the species. It is noted that "avoidance" remains the variable name for consistency with the metric naming developed in [14], but the movement patterns are simply referred to as hyper or hypo-activity to minimise confusion. Control and spike-exposed individuals were processed identically in this described calculation of $\mu(t_j)$:

$$\mu_{km}(t_j) = \frac{1}{N}\sum_{i=1}^{N} X_{kmi}(t_j) \ , \qquad t_j \in \mathrm{T}, \tag{1}$$

the multidimensional codomain is threefold, donated by $k$=1-3, corresponding to each distinct invertebrate species, giving $m$ multivariate time series observations of the $M$=22 experimental observations ($\mu_{km}(t_j)$ and $\alpha_{km}(t_j)$). Control samples serve as a baseline to ensure observed variations reflect variation and not experimental noise. Prior to B-Spline approximation, data was smoothed via Gaussian mean averaging, and for multidimensional FDA, scaling was performed from the species specific upper quartiles ($Q_{0.75}$) of the distinct locomotion distributions for each species.

## 2.3 Functional Data Analysis

Each time series data observation $\mu_k(t)$ is converted to a functional basis using cubic B-spline basis functions denoted $B_l(t)$, as described:

$$\mu_k(t) = \sum_{l=1}^{L} c_{kl} \cdot B_l(t)\,, \tag{2}$$

where $l$ represents the index of each basis function and $c_{kl}$ is the associated basis coefficient matrix (in this study separated by knots at t = [0,10,20,30,60,90,120] minutes). For functional principal component analysis in the case of univariate data (fPCA of individual species), the eigenvalue problem is solved as described,

$$C(t,s) = \frac{1}{M}\sum_{m=1}^{M} Y_m(t) \cdot Y_m(s)\,, \qquad Y_m(t) = X_m(t) - \mu_m \tag{3}$$

where each observation $X_m(t)$ is first centred by subtracting the functional mean as in (1) to give $Y_m(t)$. $C(t,s)$ is the covariance function describing covariance between two time instances $t$ and $s$ (where t and s represent any two points in T), used in the integral problem described, to calculate the Eigen functions $\phi_p(t)$ and Eigen values $\lambda_p$:

$$\int_0^T C(t,s)\phi_p(t)dt = \lambda_p\phi_p(s), \tag{4}$$

which describe the functional principal components (fPCs) calculated in (4). The fPC coefficients $a_{mp}$ are the scores of observations summed on each of the Eigen functions denoted in (5). These fPC scores described:

$$Y_m(t) = \sum_{p=1}^{P} \mathrm{a}_{mp}\phi_p(t)\,, \tag{5}$$

$$X_i(t) \approx \mu(t) + \sum_{p=1}^{p'} \mathrm{a}_{ik}\phi_{km}(t)\,, \qquad p' < M, \tag{6}$$

are used for interpretation of individual points and outliers (later visualised in various fPCA scoreplots). Similarly to standard PCA, the $p'$ number of required fPCs is determined to satisfy (6) to a chosen threshold of explained functional variance. Throughout this paper, only two fPCs are required for explained functional variance > 0.8, and thus $p'$=2, leaving two principal components: fPC1 and fPC2. The functional data packages and machine learning packages used throughout this study are the python package *scikit-fda* and *scikit-learn,*[24] as well as the R package *fda* and *mfpca*.[34,35] The dataset for mean and avoidance data is available online.[36] K-means clustering was applied to the fPCA scores derived from the behavioural responses to field alert events, enabling the classification of points into distinct clusters based on similarity in their functional profiles (the number of clusters was determined based on analysis using both the elbow method and the silhouette plot). In the case of laboratory assays, points are regrouped using bounding ellipse functions as the chemicals are known, estimated through Khachiyan optimisation.[37]

## 2.4 Chemical Analysis of Grab Samples

The raw data obtained during chemical analysis were processed to identify 803 target molecules specified by the method as detailed in the Supporting Information, carried out by LODIAG Company in France, by means of suspect screening LC-MSMS analysis. Of the 803 molecules analysed, 8 categories of chemicals were searched for: pharmaceuticals, pesticides, narcotics and their respective metabolites categories. Grouped substances in the same chemical use category are simply ordered alphabetically in later sections, where data is equally available on line.[36] For each chemical, a concentration estimate fell in one of five pre-defined categories, which are described in Table S1. Simply for visualisation purposes, retained concentrations in Table S1 are used to evaluate chemical presence in a semi-quantitative manner. The exact concentrations are presented in the Supporting Information.

## 3. Results

### 3.1 Behavioural Response for Multiple Invertebrate Species

When comparing locomotor response between the three species (*E. testacea*, *G. fossarum* and *R. auricularia*), response to some chemicals was ubiquitous across species, whilst other chemicals triggered species specific sensitivity. Copper for example triggered immediate response for all macroinvertebrate species (**Figure 1**). This shows that quasi-instantaneous response can be observed at relevant effluent concentrations for a range of invertebrate phyla (in this case 100µg$L^{-1}$). The full range of responses for the other compounds, as seen for Copper in **Figure 1**, are presented in Appendix C of the Supporting Information.

For the other compounds, an interesting observation is the immediate locomotor inhibition seen for *E. testacea* in response to methomyl (**Figure 2**A). Visual inspection of the video files confirmed no mortality and confirmed hypoactivity (locomotor inhibition), whereby *E. testacea* curls up into a ball shape (Appendix B of the Supporting Information). Overall, locomotor activity changes occurred anywhere between 1 minute post-spike (in the case of Cu response) to several hours post-spike (depending on compound and species), occasionally with no notable response at the chosen concentrations (such as *E. testacea* in response to Zn exposure at 300µg$L^{-1}$ seen in **Figure 2**). It thus appears that locomotor response dynamics are repeatable between assays, and exhibit various patterns – rapid, slow, transitory, induction or inhibition of locomotor activity) that differ between contaminants and between species.

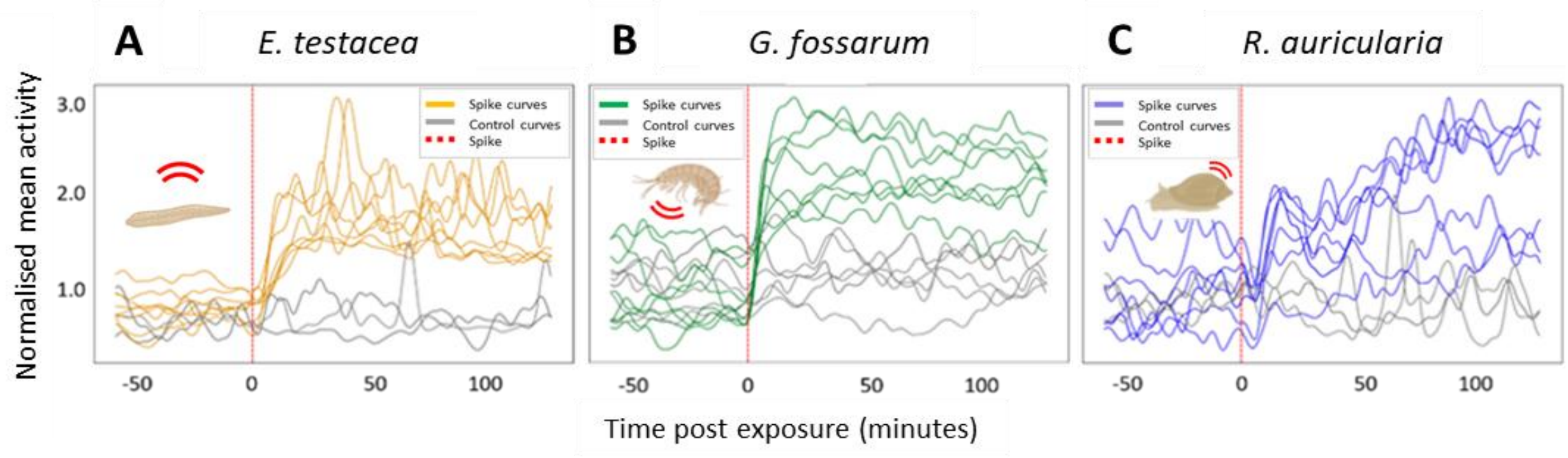

***Figure 1*** *Copper mean locomotor response for 8 spike repetitions of Copper at 100 µg$L^{-1}$, shown for (A) E. testacea, (B) G. fossarum and (C) R. auricularia (each organism form shown with associated figures). Each curve is the mean of the 16 individual activity curves (µ(t)) recorded for each Cu assay repetition.*

## 3.2 Functional Data Modelling of Single Species Responses

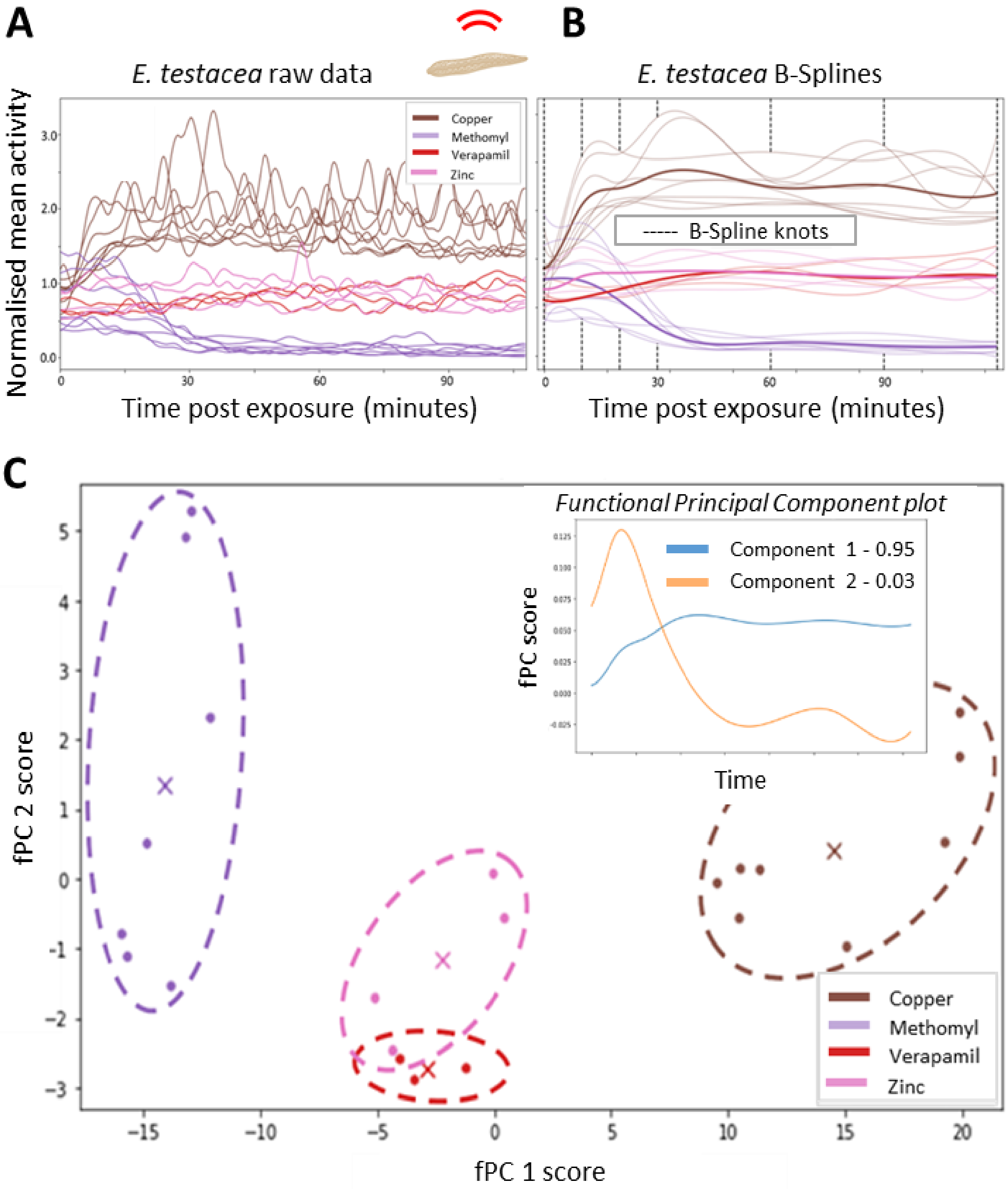

*__Figure 2__ (A) Raw data of behavioural mean response curves of E. testacea to the four distinct micropollutants (one mean curve per independent repetition). (B) The same time series data in its smoothed functional basis form (using cubic B-Splines), with nodes for B-Spline approximation marked by dashed vertical intersects. (C) The fPCA scoreplot for E.testacea; where each point represents one observation in (B). Centres marked in the bounding ellipses are used to define the approximate behavioural fingerprints (superposed bold curves in (B)). The scale of variability is not consistent between fPCs for readability. The overlaid plot (top right C) illustrates the two principal functions derived from the functional principal component analysis (fPCA). Each point is related to these functions by the extent of projection onto the fPCs, which reflects the degree of correlation or contribution these principal functions make to the behavioural patterns observed across measurements.*

FDA interpolation of locomotion curves demonstrated efficient functional approximations of the individual curves, as illustrated for *E. testacea* in passage from **Figure 2**A to **Figure 2**B, with optimal fitting using cubic order B-splines. Shorter node intervals in the first 30 minutes post spike exposure permitted better smoothing and enhanced recreation of sudden changes (**Erreur ! Source du renvoi**

**introuvable.**B). Through unsupervised exploratory data analysis, FDA reveals behavioural fingerprints in *E. testacea* in the grouped datapoints shown in the fPCA scoreplot in **Figure 2**C. Furthermore, the overlaid fPCs **Figure 2**C provide insight into temporal dynamics that characterise each cluster, so-called behavioural fingerprints. Formed through Eigen decomposition, the fPCs for *E.testacea* show that the first fPC captures the largely dominant functional mode of variation in the locomotor response, explaining 95% of the temporal variance from the mean response curve. This dominant behavioural mode is seen to be characterised by initial positive gradient over the first 30 minutes of the observation window; followed by sustained high movement amplitude. In this context, fPC1 can be interpreted as representing the general intensity of the locomotor response, while fPC2 captures more subtle species-specific variation. The fPCA scoreplot (**Figure 2**C) depicts the projection of each time series observation onto the first two fPCs. This gives a visual representation of the similarities and differences in the behavioural responses to the four micropollutant spike exposures. For Cu, the scoreplot indicates induced hyperactivity that peaks after 30 minutes and is sustained thereafter. This hyperactivity is reflected in a positive correlation with the dominant behavioural mode, as shown in the scoreplot to the right of the fPC1 axis. In contrast, for Methomyl, the behavioural fingerprint is characterised by an inverse correlation with the dominant mode, aligning with the discussed hypoactivity. The consistent negative projection in **Figure 2**C illustrates this.

Univariate behavioural fingerprints, shown for *E. testacea* in **Figure 2** and *G. fossarum* and *R. auricularia* in Figure S1, are therefore defined through functional principal components and confined by their bounding ellipse functions.[37] Fingerprint cluster centres were projected to visualise each fingerprint as a behavioural response; shown for *E. testactea* in **Figure 2**B. For *G. fossarum and E. testacea* each fingerprint gives the characteristic profile of the distinct micropollutant response, whilst for *R. auricularia* the only distinct fingerprint is for Cu (Figure S6). Overall, there is a high degree of separation between Cu and methomyl response, particularly in *E. testacea*. Explained variance of the first Eigen function (fPC1) is consistently high, showing a dominant behavioural mode characterising fingerprints for the three species.

## 3.3 Multivariate FDA – a Multispecies approach

In addition to univariate behavioural fingerprinting, we evaluated expanding the analytical scope to multivariate FDA (mFDA) to account for parallel observations across the three species. As expected from the multidimensionality that comes from the multispecies response, the principal component decomposition with mean locomotor response reveals more distinct clusters of data that corroborate the fingerprints observed for single species, with the same underlying dominant behaviour mode (**Figure 3**A). To go further, the same framework was applied to the avoidance metric, as developed in Ruck et al (2023),[14] in the multivariate case (univariate fPCA and fingerprints are shown in Appendix D of the Supporting Information for the avoidance metric). The clear separation reinforcing the multispecies metric for Figure 3A is not as clear in the avoidance metric, as the fPC 2 differences are insignificant. Nevertheless, given that the avoidance metric has been shown to be more robust in field conditions,[14] this an encouraging result for on-site behavioural fingerprinting in effluent environments, and the avoidance metric is used for the following analysis of real-time alerts.

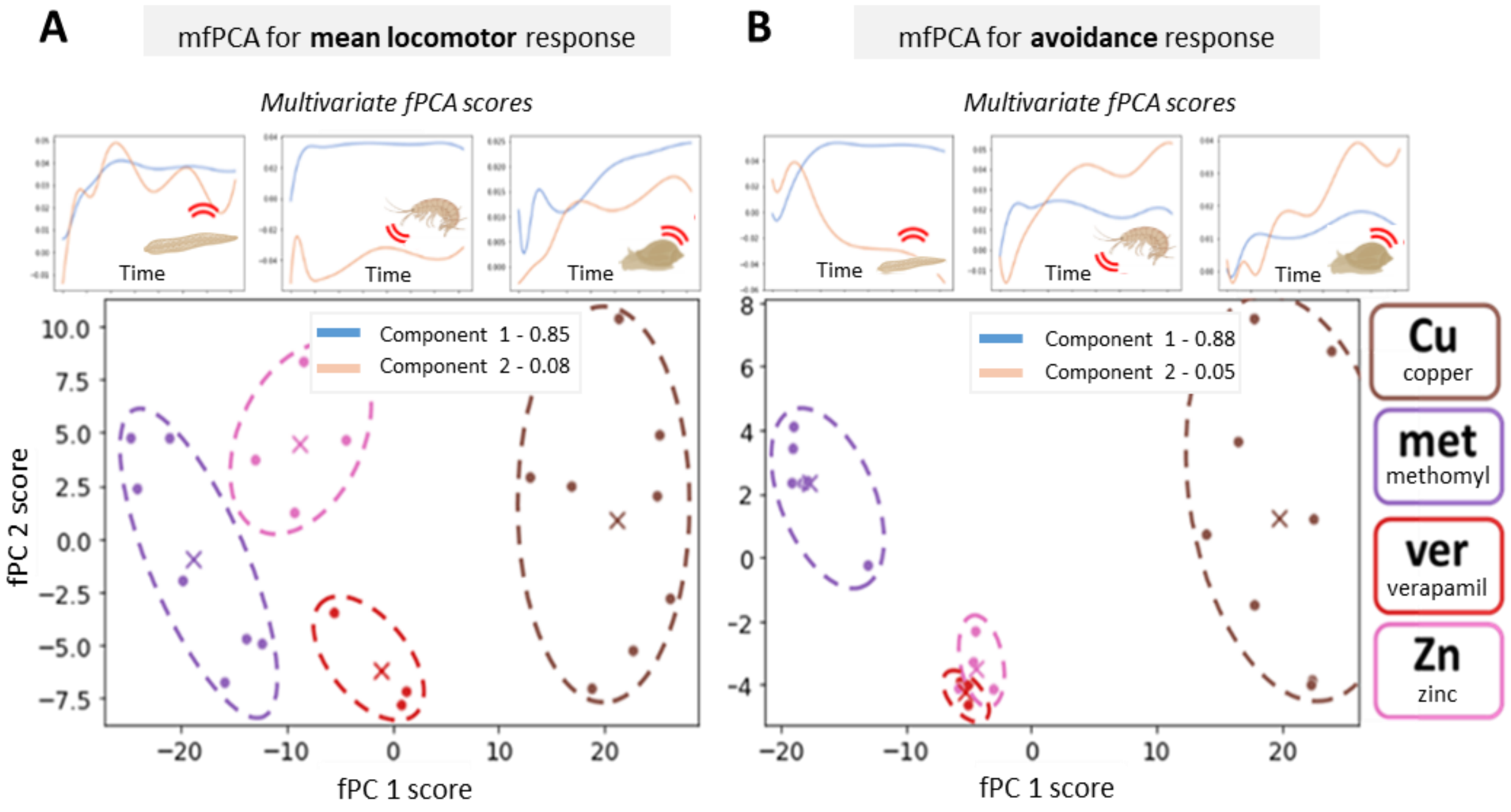

*__Figure 3__ Multidimensional fPCA for all species for (A) mean locomotor response and (B) avoidance response. The three fPC plots above each fPCA scoreplot show the species contribution to the first two multivariate fPCs. The scale of variability is not consistent between fPCs for readability.*

### 3.4 Field Study functional clustering

At the field site in north western France, operational since January 2022, 14 micropollutant surge events were accompanied by automatic grab samples over two month-long periods at the WWTP effluent (similarly to observed micropollutant events in our previous study).[14] Overall explained variance decreased to 0.79 with two fPCs (**Erreur ! Source du renvoi introuvable.**), when compared to lab tests in **Figure 3**B. This perhaps reflects the wider array of micropollutants in more complex water medium and micropollutant mixtures. Nevertheless, within the data, unique fingerprints are also observed at the effluent, grouped here with 4 k-means clusters. Interestingly, when analysing dates associated with grouped clusters, similarity in locomotor profiles is shown between alerts up to 6 months apart, although a global separation is noted for both the chemical nature and behavioural profile of events sampled between the two periods (as detailed in the Supporting Information). Appendix E of the Supporting Information presents the locomotor responses for each point in **Figure 4**A, along with the corresponding chemical profiles, as detailed for the four highlighted samples in **Figure 4**B. Determining a definitive global seasonal trend would require longitudinal data over multiple years. **Figure 4**B shows that when nearby points in common behavioural fingerprints are compared, the chemical profiles are strikingly similar. Furthermore, the control sample (collected outside of any behavioural alert) is the only sample to show minimal to no concentration for all dosed chemicals.

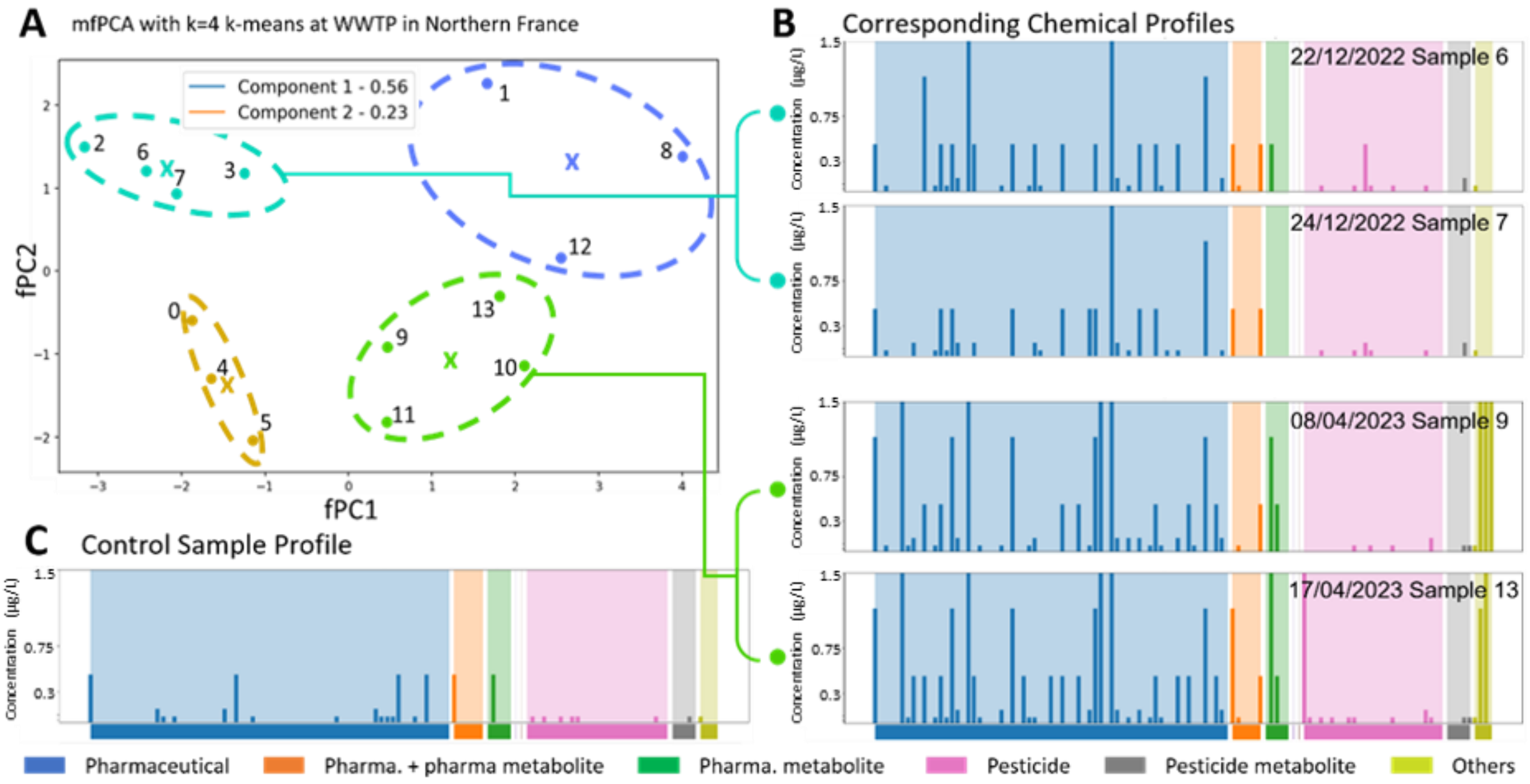

*__Figure 4__ mfPCA of the behavioural alerts accompanied by grab samples for chemical screening analysis. (A) The mfPCA shows 4 k-means defined fingerprints, describing the functional variability within the alert profiles uniquely for the behavioural similarity. (B) Shows the chemical screening profiles of four grab samples collected during four behavioural alerts falling into two distinct fingerprints (cyan and green clusters 2 and 4); semi-quantitative abundance of 803 chemicals is shown with barplots, grouped by chemical categories with their corresponding groups below. (C) Shows a control sample in the absence of avoidance. The ensemble of the chemical samples is shown in the Supporting Information.*

## 4. Discussion

### 4.1 Biological findings and Sensitivity to 3 Classes of Micropollutants

The first significant finding was the successful extension of our developed biotest to two new species, where we saw clear locomotor response to three classes of micropollutants. With regards to inter-species variability, the observation of hypoactivity in *E. testacea* in response to methomyl is particularly interesting. This is coherent with various findings for locomotor inhibition in behavioural exposure studies, especially for similar anticholinesterase neurotoxic insecticides in aquatic vertebrate larvae.[38] This form of bioactivity response may be referred to as a stress-induced defensive posture, which may either serve as a protective mechanism to reduce vulnerability or a posture to induce drift like avoidance.[39-41] Previously, we assumed that carefully conditioning towards a minimal activity state of organism movement may only channel bioactivity through hyperactivity.[14] It is thus shown that hypoactivity can also be captured through this bioassay in response to certain chemicals.

Locomotor response is observed not only for insecticides, but also for a pharmaceutical and two commonly used metals. The concentration levels used in laboratory assays to detect relevant changes in organism locomotor activity may raise questions about the method's sensitivity in relation to environmental realism. Compilations of available data for French urban wastewater treatment plant effluents[42] demonstrate that average concentrations of many trace metals exceed µg/L levels (e.g., copper: 4 µg/L, zinc: 35 µg/L), a pattern also observed for various pharmaceuticals and pesticides. Furthermore, contaminant surge events in effluents can result in concentrations significantly higher than these averages. For copper and methomyl, the concentrations tested in the laboratory align with maximum concentrations reported in certain effluent monitoring campaigns.[14] In the French dataset of WWTP effluent monitoring,[42] maximum measured concentrations of micropollutants are approximately ten times higher than median values. These ten-fold ratios are likely underestimates, as traditional effluent grab sampling often misses surge concentrations. In industrial effluents, concentrations of specific contaminants can be much higher than those in urban wastewater effluents. Additionally, studies indicate that contaminant cocktails present in real effluents can exhibit additive toxic effects. Finally, the relevance of behavioural responses observed in mono-substance laboratory tests to real effluent monitoring is supported by our field study (Figure 4), where detected bioactivity is similar in magnitude to laboratory responses.

The distinct fingerprints for the different classes of micropollutant adds further evidence that the multi-species biotest as a good approach for non-targeted micropollutant detection, sensitive to a range of compound types at sublethal concentrations relevant for environmental monitoring.[2,14,43] On top of previously tested classes, the behavioural distinction for locomotor response to a commonly found pharmaceutical is promising, as these are a common class of CEC shown to routinely bypass elimination during secondary WWTP treatment.[32] Interestingly, the two micropollutants closest in the multi-species functional space (**Figure 3**) were not the two metals Cu and Zn, but rather Zn and the pharmaceutical verapamil. The clear difference in the signature of Cu and Zn, shows that fingerprinting goes further than simply classifying chemicals by micropollutant type. As expected, other factors must explain the complex interactions that govern behavioural response. These merit further investigation, such as targeted biological pathways that could be revealed to trigger certain behavioural phenotypes. The variety in the locomotor response patterns across species suggests that for a larger number of micropollutants, fingerprinting would not be limited to only a handful of profiles. A complementary open question for the broader practical application of this method concerns the extent to which behavioural response patterns are transferable between phylogenetically related species. Are there taxonomically equivalent species that could be used interchangeably? Behavioural ecotoxicology has already shown that this may raise challenging scientific questions, as behavioural differences can exist even between closely related species (see [44] for a recent findings in gammarids) or even more subtle changes between different populations of the same species.[45]

## 4.2 FDA implementation for Behavioural Fingerprinting

In the presented functional framework, response curves are reduced to finite points in space through fPCA, which provides a framework for clear behavioural fingerprinting. This allows for a nuanced understanding of the temporal patterns in invertebrate responses to different micropollutants. The fPCs highlight common and dominant modes of behaviour, while the scoreplot facilitates the comparison of individual trajectories, revealing variations and correlations with the identified intrinsic behavioural modes captured. Indeed, fingerprinting exists in many different fields, such as in audio fingerprinting for extract recognition,[40,41] weather pattern identification,[43] and computational analysis of biological behaviour.[17,46-48] There are common themes in the machine learning clustering methods used to define fingerprints, however something that appears more common in behavioural analysis and research, particularly in ecotoxicology, is the use of overly simplistic temporal descriptors for clustering, such as those defined for hierarchical clustering.[17] Simple descriptors such as number of bouts of activity and inactivity for example, as highlighted by Egnor and Branson,[46] often face significant limitations. Beyond the bias associated with predefining the metrics, it is usually an oversimplification of complex interactions that discards temporal information, losing nuance in behavioural dynamics.

Our proposed framework shows high interpretability and flexibility, with the use of irregular knot spacing for optimised data smoothing, and even the ability to combine non-uniform data structures in mFDA if necessary (images and univariate data for example).[35] FDA should be considered more often for interpreting functional like data in ecotoxicology and time series data in biological sciences. A recent issue dedicated to Data Science in environmental science showcased papers from the last seven years of relevant contributions in the field.[46] Critical reviews stressed the associated difficulty with models lacking in interpretability,[47-48] however despite over hundreds of cited machine learning keywords in one review,[49] there was no mention of FDA despite various other papers that would be well suited to the framework.[50,51] Ironically, FDA's departure from conventional multivariate statistics, treating observations as continuous functions, aligns well with the chronic and sequential nature of biological behavioural tests used in typical ecotoxicology, providing a framework that merits further attention in the field.

Further testing of micropollutants, as well as the extensive data collection from field monitoring, may open perspectives for more complex machine learning FDA applications. For example, mean response points display greater spread along the second fPC in the multivariate approach (**Figure 3**), showing that this component could capture subtle differences that could be exploited through supervised approaches in the future. It helps to think of the first functional principal component (fPC1) as capturing the most dominant shape of the behavioural response. For example, in *E. testacea*, fPC1 shows an increasing activity pattern in **Figure 3**. When we look at the actual corresponding data in **Figure 2**, we see that copper triggers just such a hyperactive response, which explains why it appears on the positive side of this plot and **Figure 3**b. Conversely, methomyl shows a drop in activity and sits on the negative side of fPC1. The second component (fPC2), while less dominant, can still help us pick up smaller differences, but in **Figure 4** where it has higher statistical significance it captures cases where one species react differently to another across the response, thus becoming a part of fingerprint we're trying to identify.

In this study, there was little interest for complex machine learning classification methods given the size of the dataset. However, with increased data points for 100 or so micropollutants, machine learning methods already exist that may help distinguish behavioural response, such as unsupervised functional clustering[25] or even supervised methods such as linear discriminant analysis.[52]

### 4.3 Outlook for Real-Time Monitoring

In our previous study, the methomyl reaction for *G. fossarum* was shown to be triggered only after several hours,[14] which contrasted the sub-half-hour reaction time of *E. testacea* observed in this study. Thanks to the incorporation of a multispecies approach, and the sometimes striking differences in the times for onset hyperactivity surge, it was possible to reduce the observation window for behavioural fingerprinting to 2-hours post the moment of spike exposure. The advantage of this in the context of real-time monitoring is the potential to characterise ongoing pollution surge events in real-time, which could revolutionise the way operational wastewater management occurs. This is encouraging for the incorporation of fingerprinting to real-time monitoring of WWTP effluents, where the idea of accompanying alerts with characterisation of the pollution nature is novel.

At WWTP effluents, where episodic pollution is common, recurring events of similarly natured pollution can be expected. This could take the form of repeated discharge from a single upstream industrial polluter or of recurring pesticide surge for example.[53] Field monitoring data taken from the north west of France was used to demonstrate the ability of an FDA framework to classify these alerts based on behavioural similarity. The chemical analysis of corresponding samples provided convincing evidence that behavioural fingerprints in WWTP effluents are made up of similar natured micropollutant mixtures that trigger alerts. If such similarities are successfully characterised, it could revolutionise the way in which operational water treatment can limit critical events of micropollutant transfer, through enhanced understanding of the passing pollutants. Whilst it is not proposed that this biotest will identify specific contaminants, let alone their concentrations, behavioural profiles could effectively describe the dynamics of surge chemical cocktail discharge of varying natures in urban and industrial effluents, even in the case of poorly characterised CECs. A further chance to improve the impact of such detection would be to monitor earlier in the WWTP process (such as after primary treatment). This work is ongoing in parallel to effluent studies, with a focus on addressing the practical challenges of operating the bypass system with semi-treated influent - particularly in terms of organism viability and system maintenance. Whether behavioural fingerprints are transferable from water matrix type to type or geographical site remains an unexplored topic in this research. While variation is expected due to site-specific pollutant mixtures, the underlying statistical methodology is transferable which is key. This will allow operators to build site-specific libraries of behavioural profiles that improve over time, perhaps with transferability between certain sites.

This study has established the relevance of a locomotor multispecies biotest as a non-targeted approach to rapid micropollutant detection, through behavioural response to four distinct micropollutants observed for three phyla of aquatic invertebrates. We have demonstrated the effectiveness of FDA as a framework to determine interpretable behavioural fingerprints of contaminants. The methodology proved well suited for grouping surge contamination events of different natures via similar behaviour profiles, extracted from on-site WWTP effluent monitoring data. In order to fully test the discrimination potential of behavioural fingerprinting within the chemical diversity of environmental contamination, one next step is to expand the testing scope for the massive testing of one hundred or so micropollutants in the laboratory. The application of the approach to dozens of active ToxMate field case studies is also relevant in the comparison of bioactivity reactions in varied effluents. As WWTPs are poorly equipped to deal with real-time micropollutant monitoring, on-line behavioural fingerprinting implementation - potentially coupled with piloted grab sampling - appears as a new solution to better characterise contamination transfers within WWTPs, and potentially develop innovative strategies to adapt real-time treatment and re-use of wastewater.

**Acknowledgment**: This research received financial support from Région Auvergne-Rhône-Alpes (projet Pack Ambition Recherche 2020 - ToxPrints), ANRT (Cifre PhD Grant of GR), and European Union (H2020-EIC-SMEInst-2018-2020 Grant#881495). Experimentation for data collection would also not have been possible without the help of A. Descamps, H. Quéau, T. Cavannah, L. Garnero, C. Grant, M. Dauphin,K, Montalbano and the Chartres metropole (Julien Pelleray and Julien Bordeaux).

## SUPPORTING INFORMATION

### *Appendix 0 – Stabilisation and Spike laboratory testing with the ToxMate for a multispecies approach*

The initial study [14] focused solely on *Gammarus fossarum,* a well-established bioindicator species. This study defined avoidance behaviour as an effective metric under controlled and field conditions to observe variation in surge micropollutant concentrations. In this study, a minimal activity model to distinguish between normal baseline behavioural and behavioural response, after careful acclimatisation period in the laboratory and ToxMate, was shown. This baseline model involved careful 2 week stabilisation in the lab priori to exposure to ToxMate conditions, as well as at least 72 hour stabilisation in the ToxMate to ensure a basal activity baseline, after which the 16 individuals in each assay showed reproducible locomotor patterns in the absence of contaminants (controls or control conditions), as well as with spikes of 4 methomyl and copper exposures. This data is recycled, along with new data for the current study.

In the initial study, the ToxMate system was not used to its full potential, as the ToxMate is designed with three independent chambers, each capable of hosting 16 individuals of a given species. The multispecies functionality is particularly valuable for increasing scope of pollutant type detection, and a major focus of this study.

In this study, the same methodology and experimental protocol were extended to two additional invertebrate species:

- *Erpobdella testacea* (a freshwater leech species described in the paper)
- *Radix auricularia* (a widespread freshwater gastropod)

These species were chosen on their distinct phylogenetic positions when compared to *Gammarus*, and allowing for broader biofunctional diversity in an aim to cover as many specie sensitivities to micropollutant families as possible (as opposed to choice based on ecotoxicological or ecological relevance, which is not of interest in the real-time alerting operational vision of the ToxMate).

**Spike Exposure Protocol**

Once the stabilisation has been carried out, a concentrated solution of the desired chemical is introduced into the ToxMate chamber as described in Figure S0-1. Each of the three chambers is linked to a pump in the basin that ensure homogenous and quasi instant mixture (as described in [14].

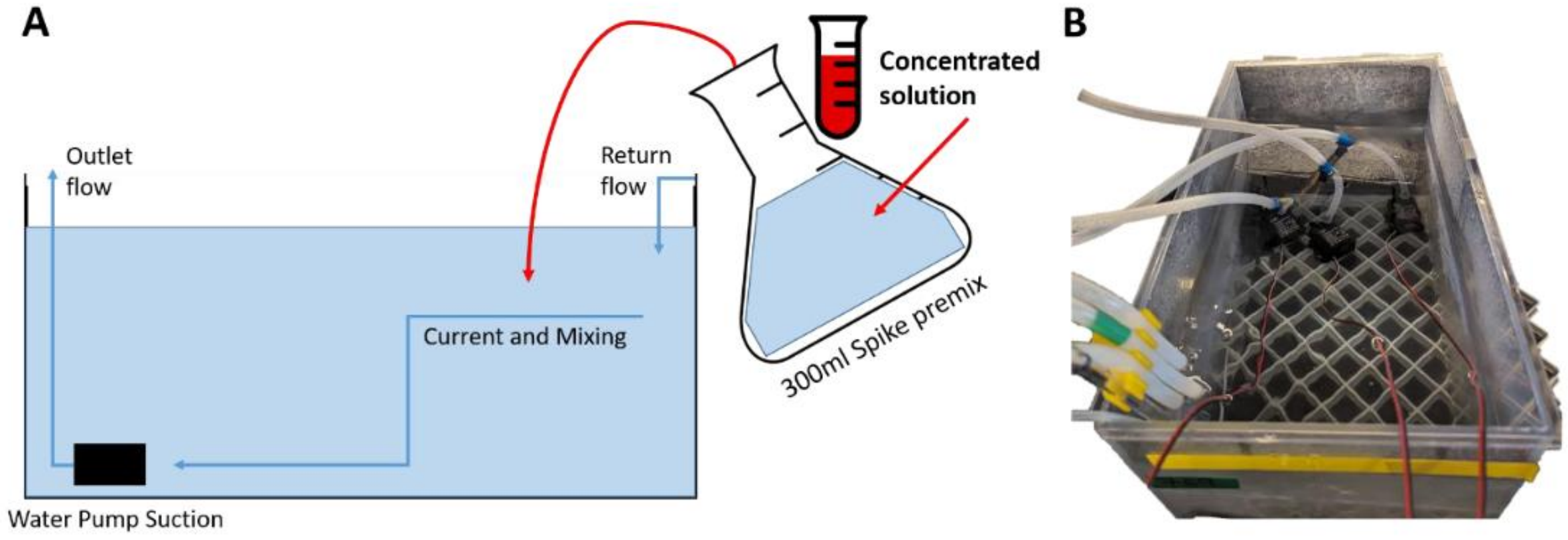

*Figure S0.1 Experimental mixing procedure. (A) Visualisation of the spike mixing in the 15L water tank of the closed circuit through the premix in a 300ml Erlenmeyer (B) The above view of an operating water tank with the three pumps aligned at the far end, and the oxygenation and return flow at the front end, where spike solutions are introduced.*

### Scaling for Cross Species Comparison

It is noted that the quantile used in the avoidance metric remains unchanged for the two additional species. To facilitate comparison of behavioural changes across the species, all individual locomotor data were linearly scaled. Scaling was based on the species specific median activity level observed during control conditions. This preserves the relative dynamics of behavioural response while allowing standardised interpretation. This is largely sufficient in this proof of concept stage, as shown by the matching distributions post scaling in FigureS0-2.

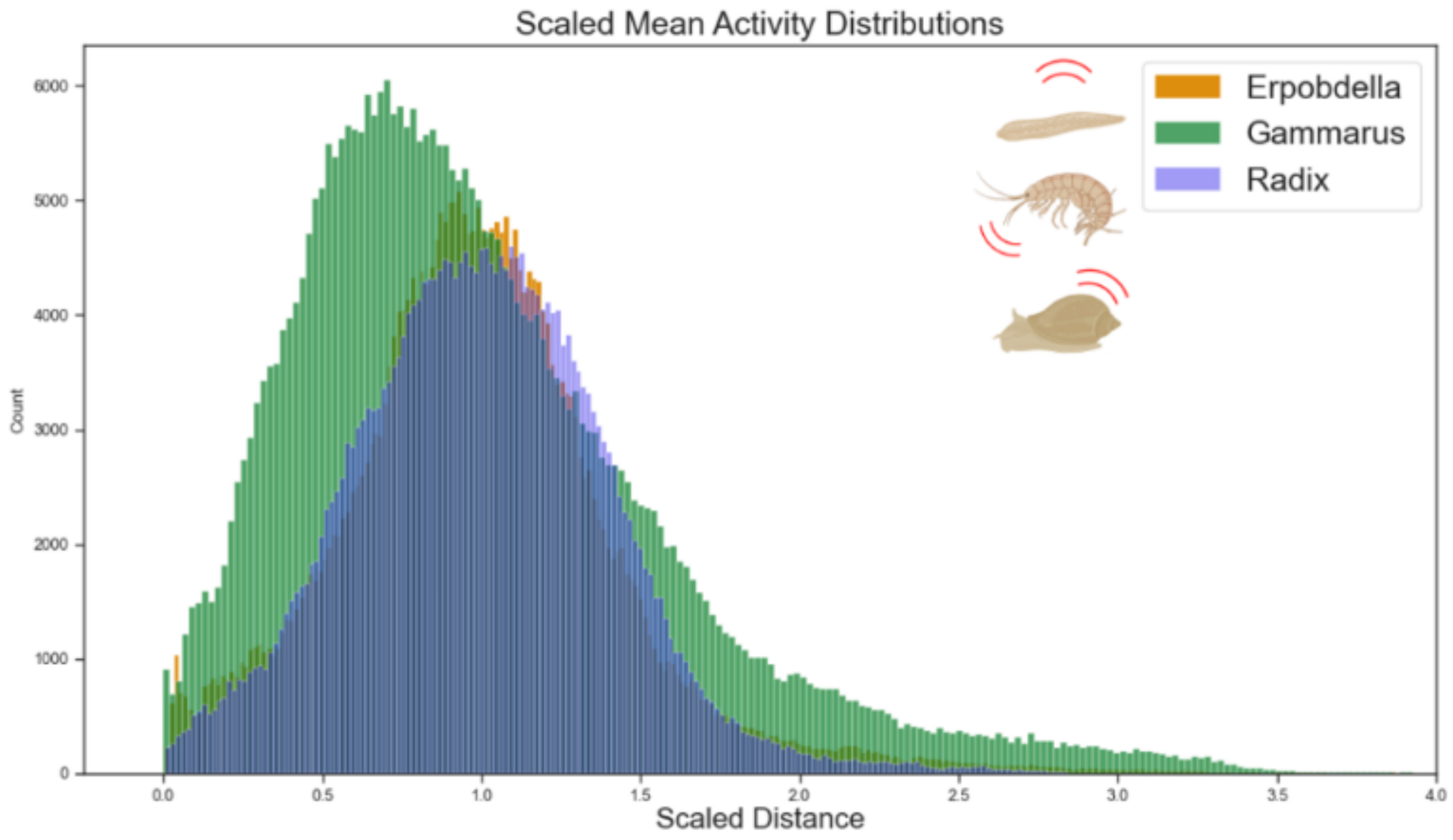

*Figure S0.2 The scaled activity distributions from data 2 hours prior to spike exposure from the ensemble of data throughout the PhD, for E. testacea, G. fossarum and R. auricularia*

## *Appendix A – Chemical Analysis*

The Table S1 details the categories used for semi-quantitative representation of chemical fingerprints in the paper.

*Table S1 Chemical concentration brackets as detailed by outsourced chemical analysis, for the 803 dosed chemicals in on site ToxMate grab samples.*

| **Concentration bracket ($\mu gL^{-1}$)** | | | | | | |
|---|---|---|---|---|---|---|
| No detection | Unquantified | <0.05 | 0.05-0.1 | 0.1-0.5 | 0.5-1.0 | >1.0 |
| **Concentration retained for visualisation (and discharge estimates) ($\mu gL^{-1}$)** | | | | | | |
| 0 | 0.01 | 0.05 | 0.1 | 0.5 | 1.0 | 1.5 |

The details of the chemical analysis are in the document provided by LODIAG company (France), in the attached Supplementary Materials.

## *Appendix B – E. testacea shows hypoactive avoidance*

The defence posture is visualised during a ToxMate experiment observation.

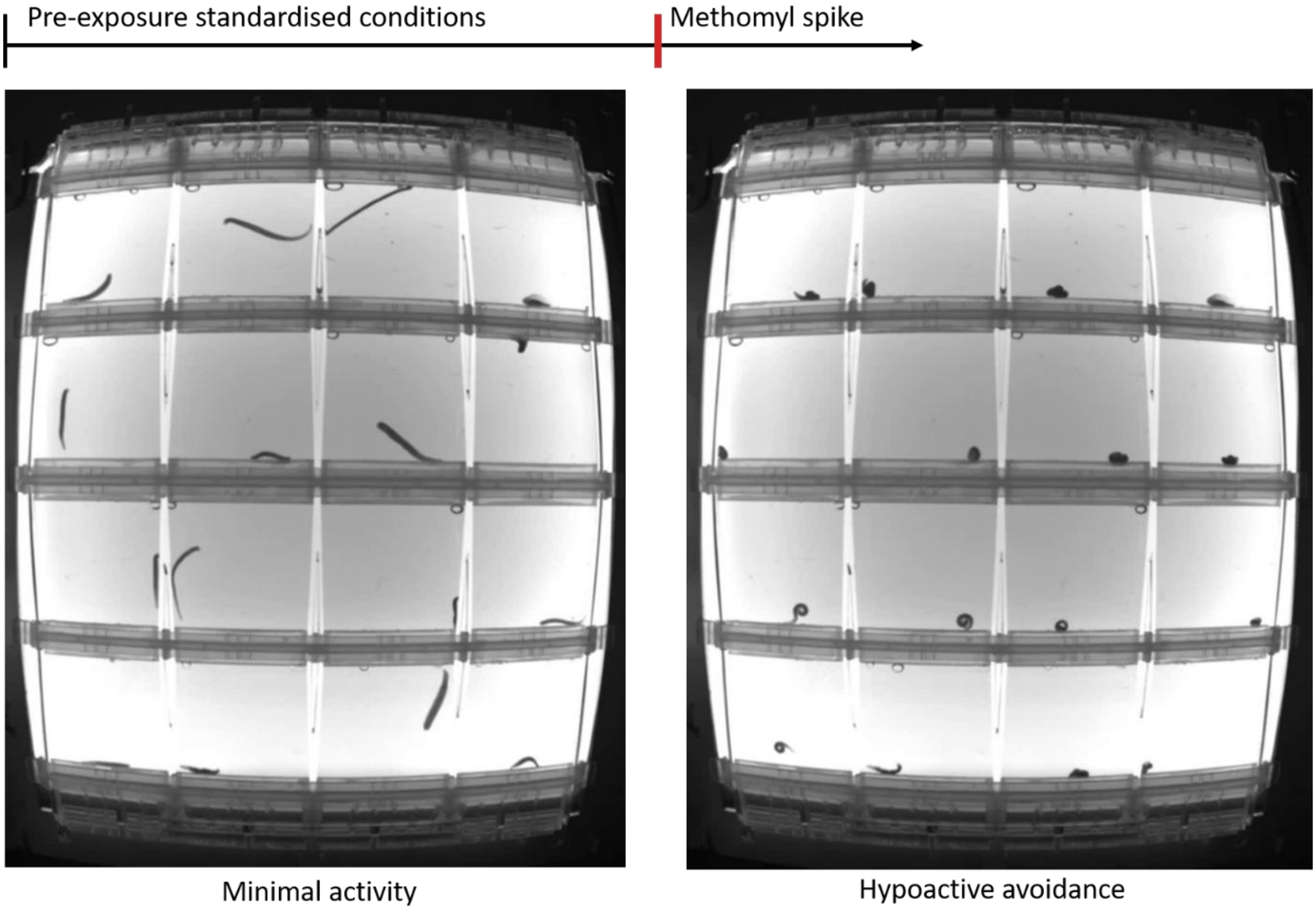

*Figure S1 Methomyl exposure at 125ug$L^{-1}$ provokes a defensive hypoactive avoidance reaction in sub 20 minutes of exposure.*

## *Appendix C – Mean and Avoidance Plots for Four Distinct Micropollutants*

Figure S2 shows the Copper observations for all species and repetitions.

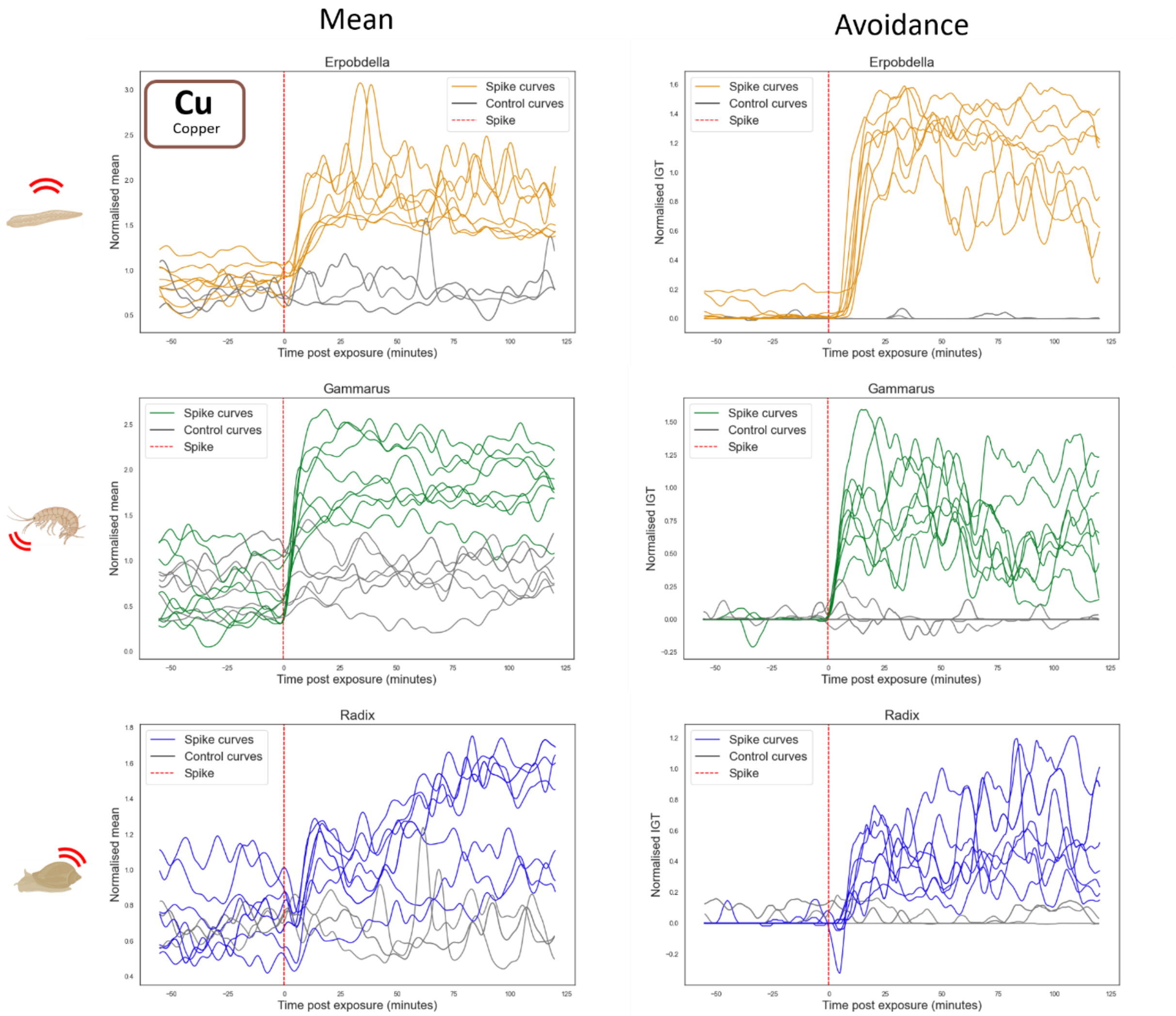

*Figure S2 Copper locomotor response for 8 observations of spike exposure at 100µgL$^{-1}$ for E. testacea (orange), G. fossarum (green), R. auricularia (blue) in comparison to control assays.*

Figure S3 shows the methomyl observations for all species and repetitions.

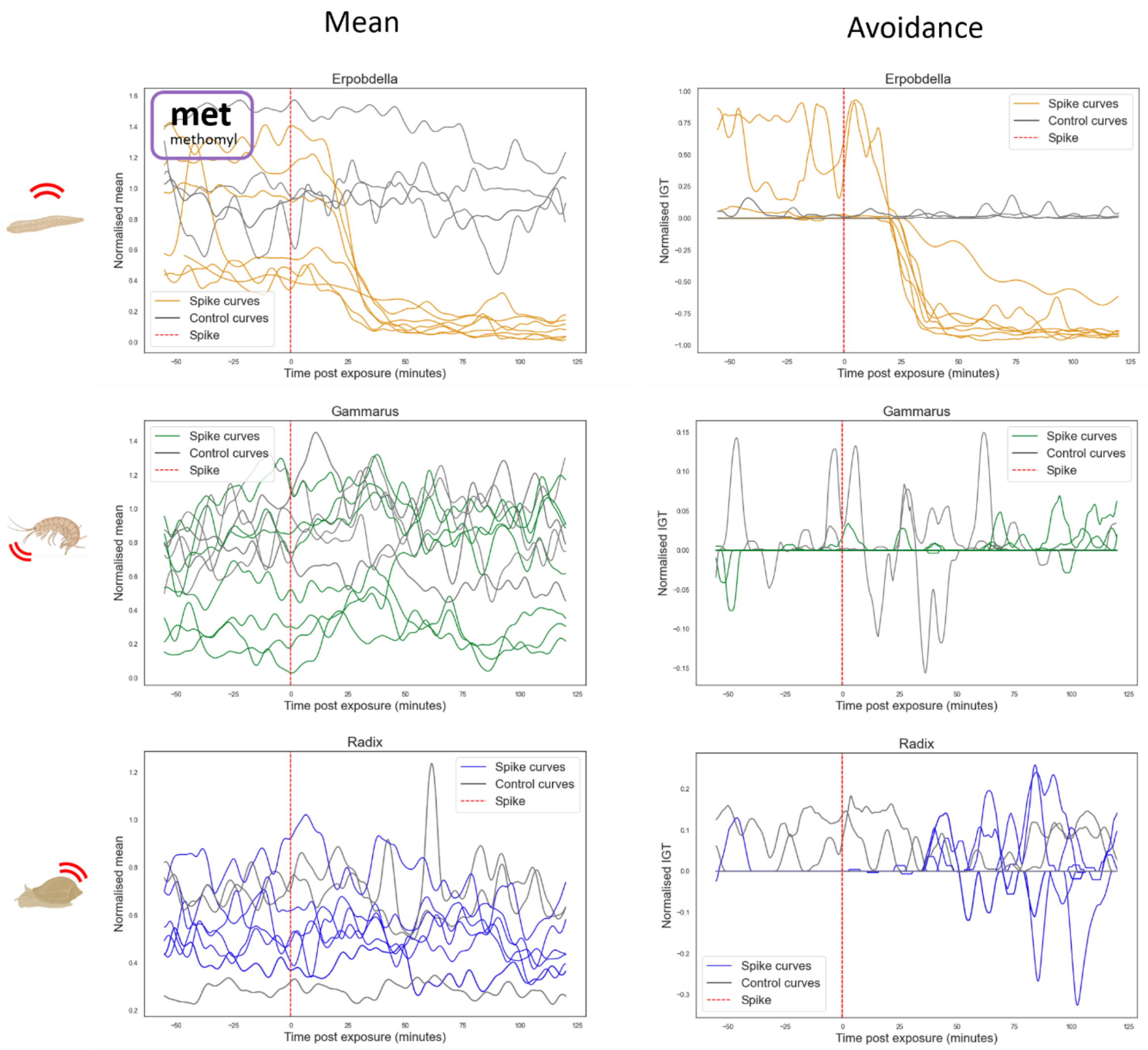

*Figure S3 Methomyl locomotor response for 7 observations of spike exposure at 125µgL-1 for E. testacea (orange), G. fossarum (green), R. auricularia (blue) in comparison to control assays.*

Figure S4 shows the verapamil observations for all species and repetitions.

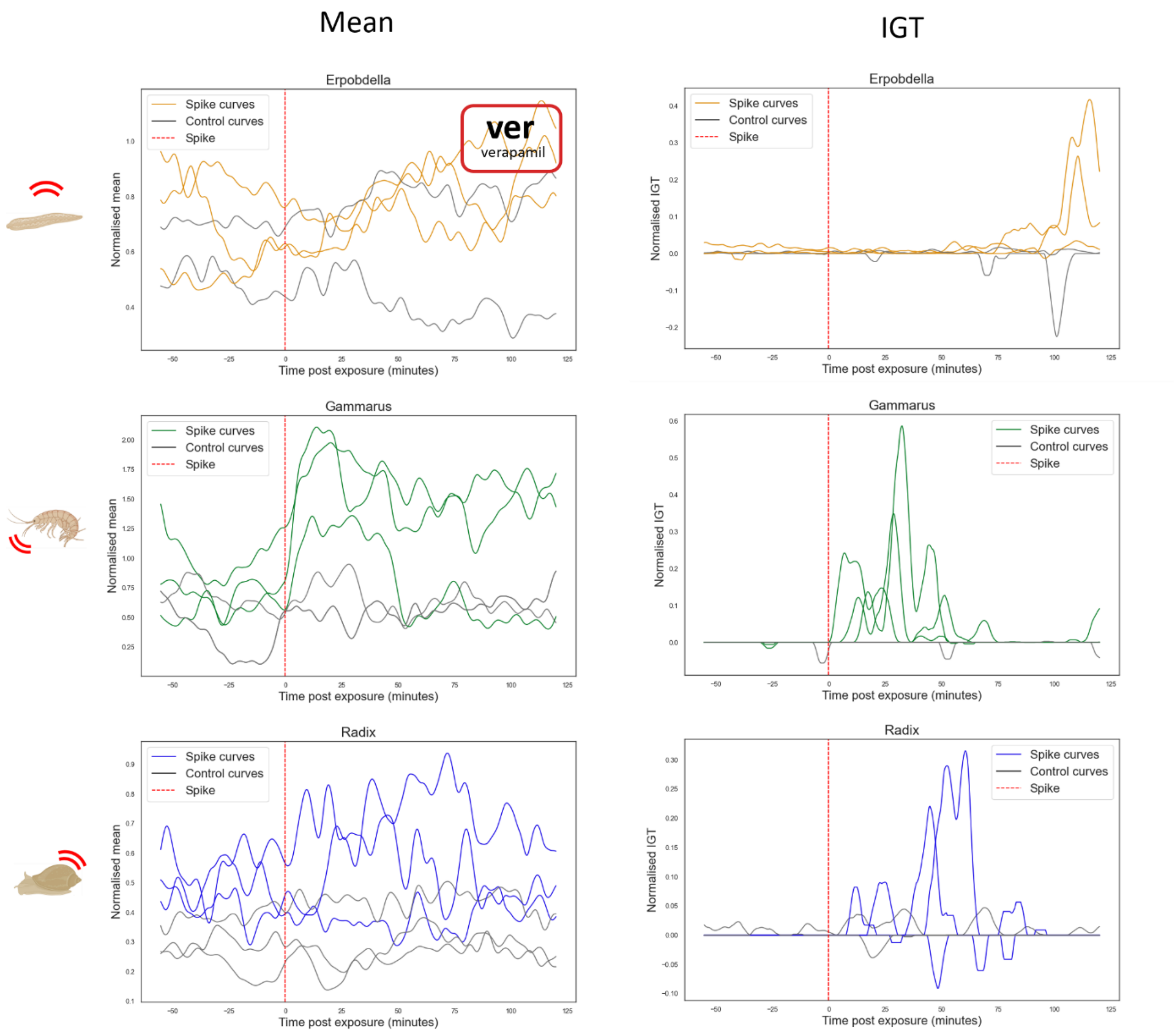

*Figure S4 Verapamil locomotor response for 3 observations of spike exposure at 120µgL⁻¹ for E. testacea (orange), G. fossarum (green), R. auricularia (blue) in comparison to control assays.*

Figure S5 shows the Zinc observations for all species and repetitions.

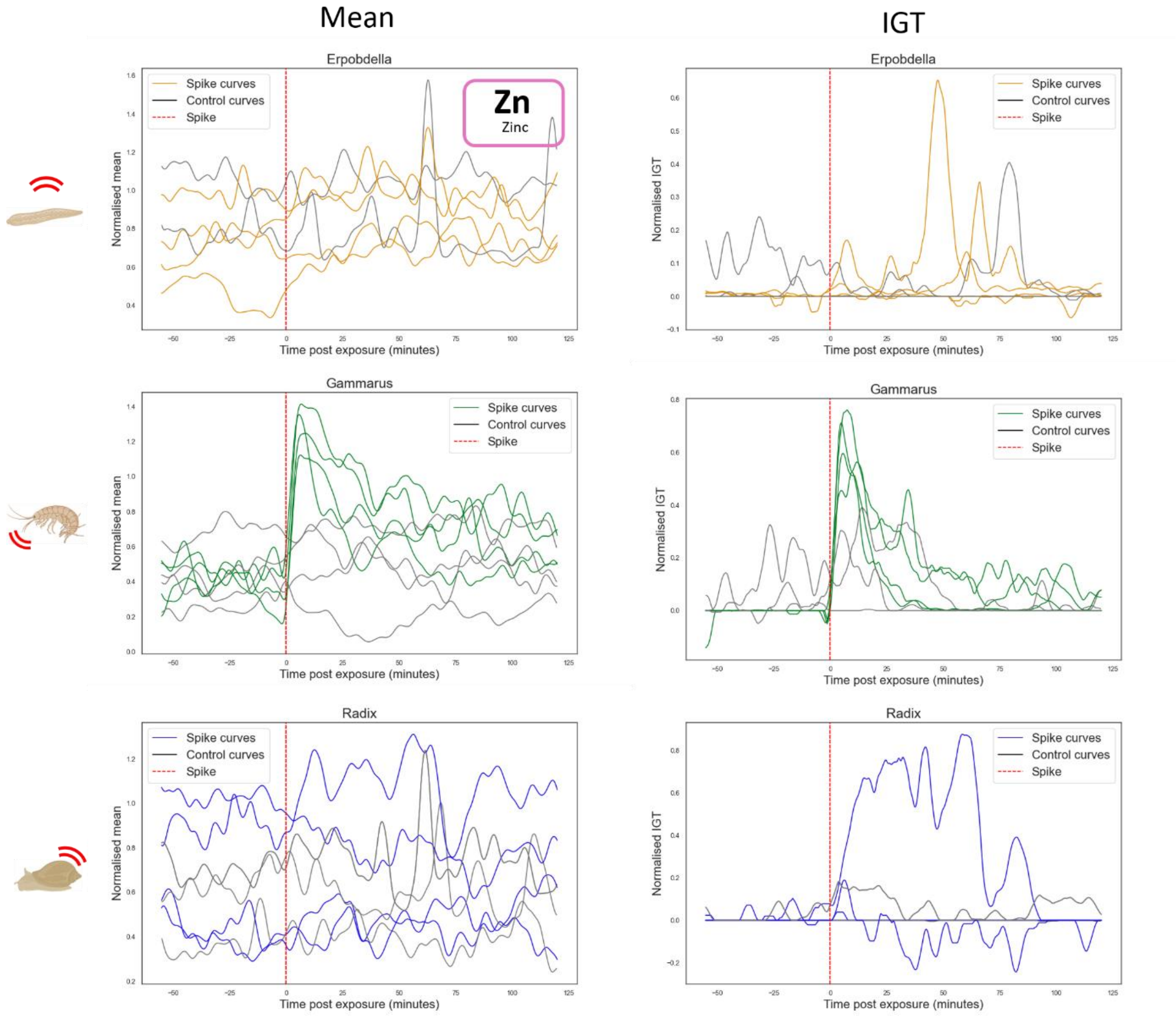

*Figure S5 Zinc locomotor response for 4 observations of spike exposure at 325µgL$^{-1}$ for E. testacea (orange), G. fossarum (green), R. auricularia (blue) in comparison to control assays.*

## *Appendix D – Univariate fPCA plots*

Figure S6 represents the mean univariate fingerprint plots for *G. fossarum* and *R. auricularia*.

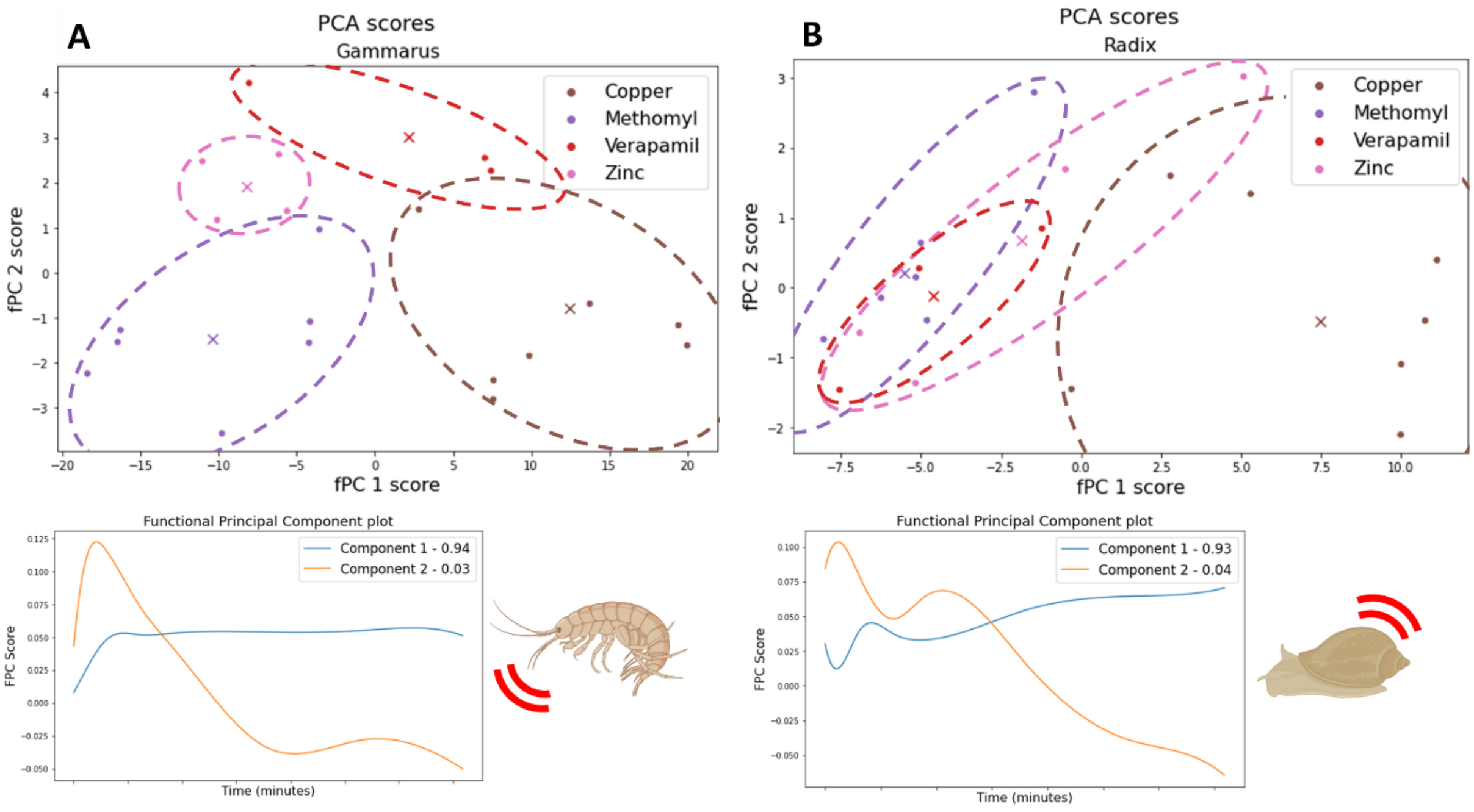

*Figure S6 fPC and fPCA plot for G. fossarum showing 95% of explained variance. (B) fPC and fPCA plot for R. auricularia showing 95% of explained variance*

The following three plots in Figure S7 represent the resulting univariate fPCA taking the avoidance signal *α(t)* as opposed to the mean functional variable shown in §3.1.

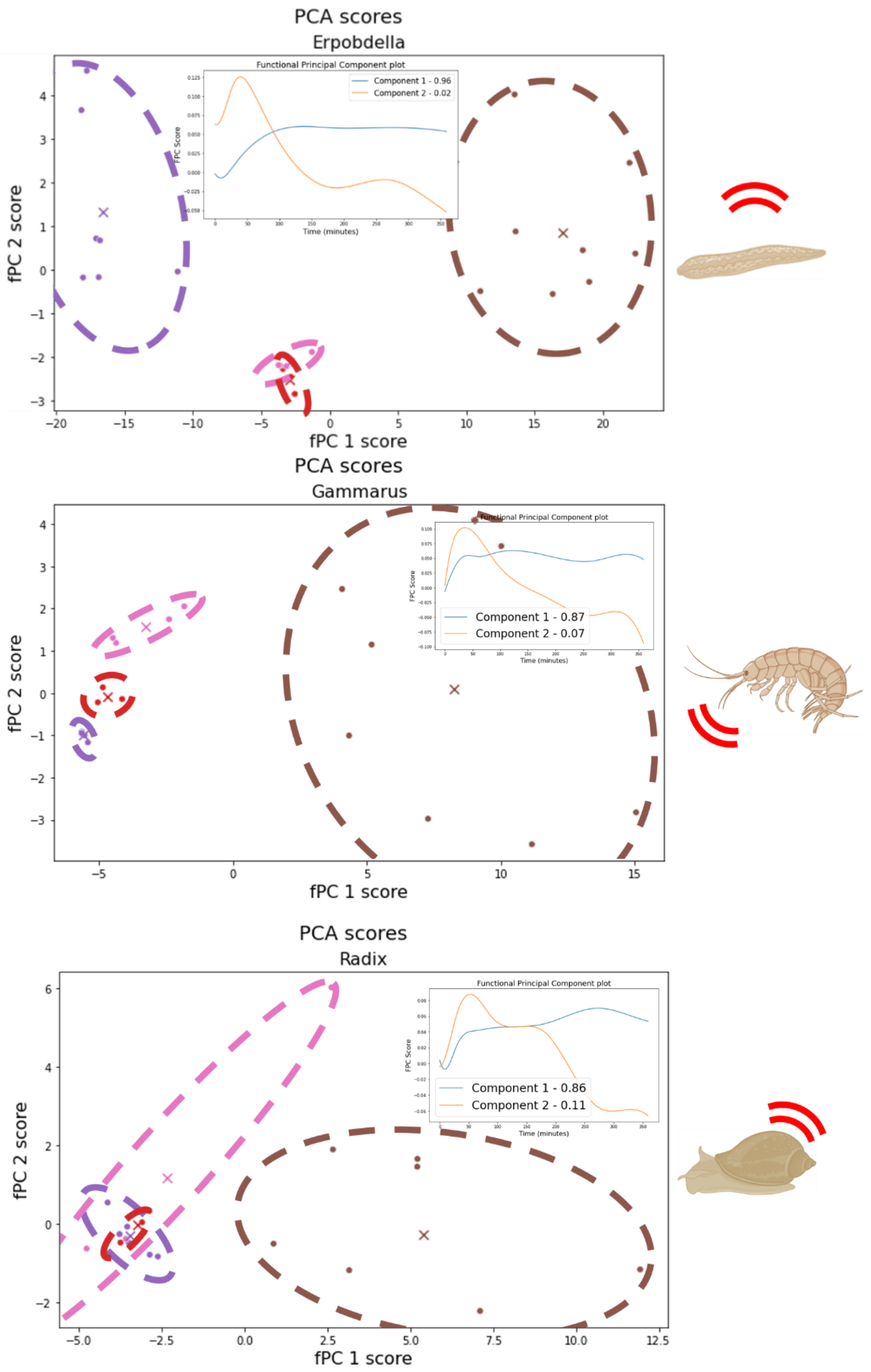

*Figure S7 fPCA scoreplot and fPCS for the avoidance signal of the three invertebrate species.*

## *Appendix E Field Data Clusters Visualised Individually*

This appendix supplies complementary details on the mfPCA performed on videotracking data during the behavioural alerts from the field WWTP monitoring (Fig4 of the main document). It serves to visualise avoidance curves behind points in Figure 4A (Figure S9-12). The fPCA figure is shown in Figure S8 with labelled data points associated to the curves in Clusters 1-4 (Figures S9-12). The upper panels of Figure S8 show the elbow plot and the silhouette plot supporting the four clusters formed. It should be noted that the selection of 4 clusters is somewhat arbitrary, as the goal is simply to demonstrate the feasibility of the developed method when treating field data. Here, the Silhouette scores are comparable for 2, 3, and 4 clusters, but decline noticeably from 5 clusters onward. Meanwhile, the inertia criterion in the elbow analysis shows clear improvement from 3 to 4 clusters. Thus, k=4 emerges as the optimal trade-off for clustering and subsequent analyses.

In the supplementary table, the LODIAG screening method targets over 800 compounds, and the subset detected in at least one sample are presented in the supplementary tables for relevance and clarity.

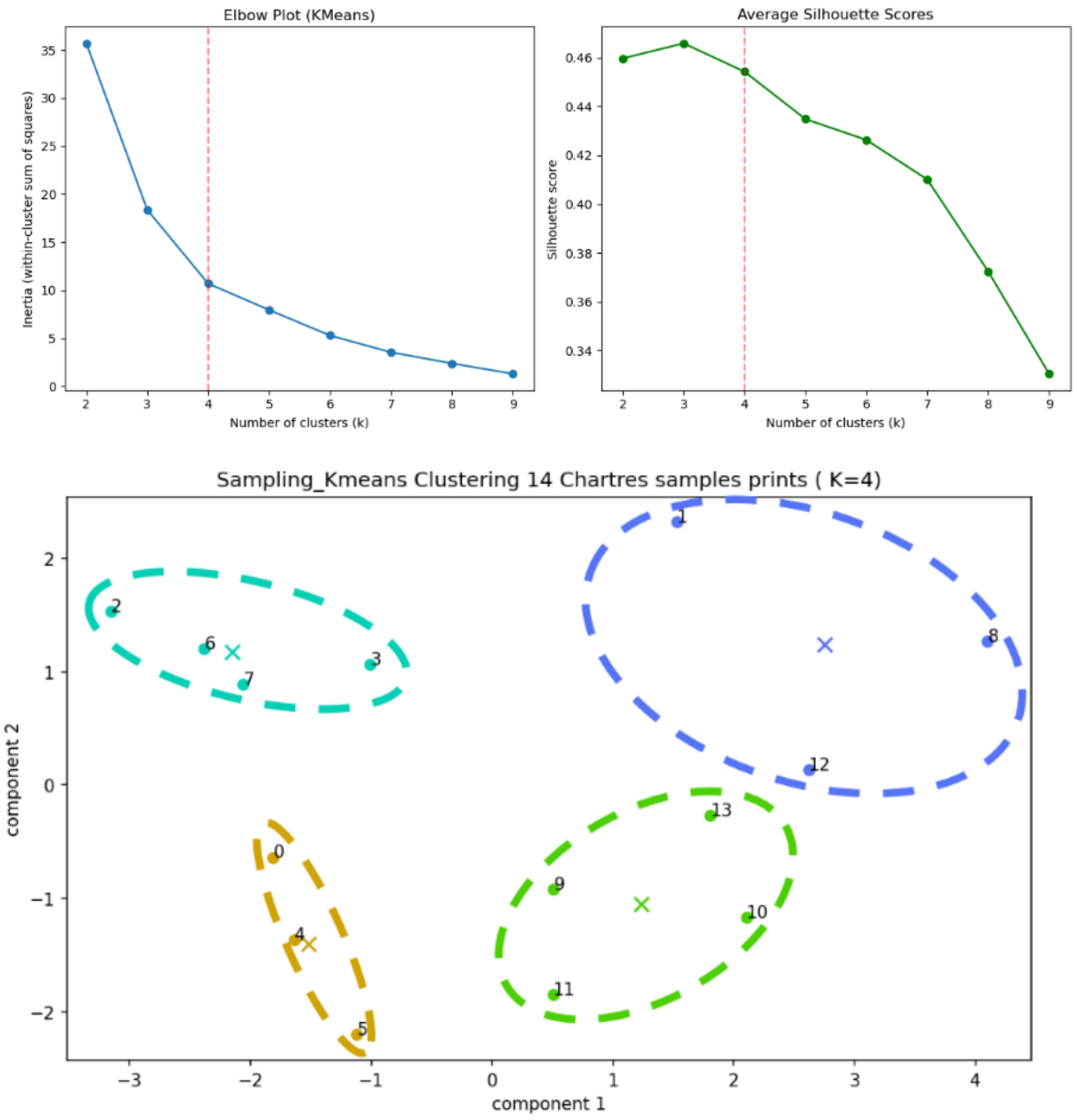

*Figure S8 Clustering of all field events detected at the monitoring site in northern France, as in Figure 4 of the main text.*

## Cluster 1

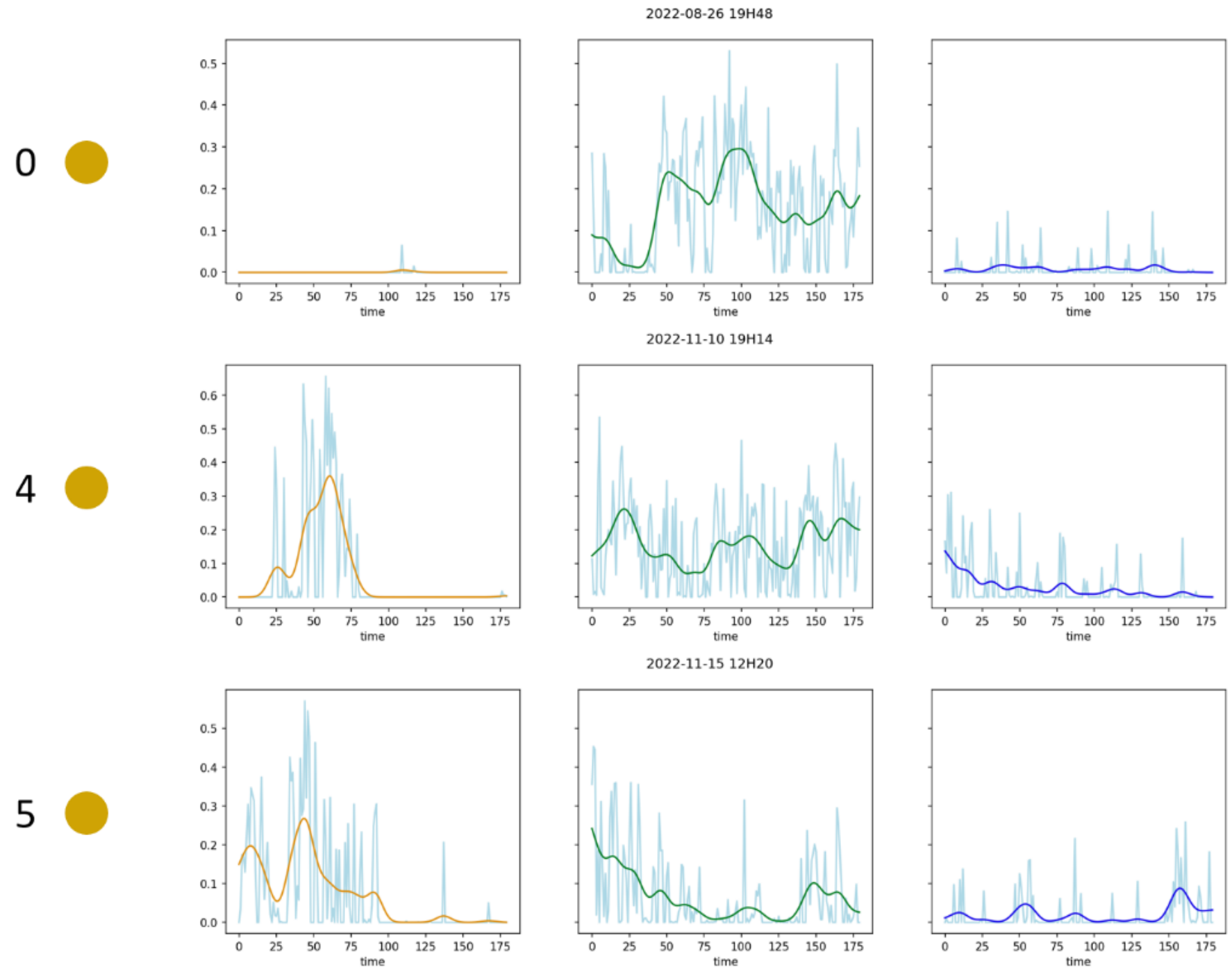

*Figure S9 Cluster 1 avoidance curves from field monitoring*

## Cluster 2

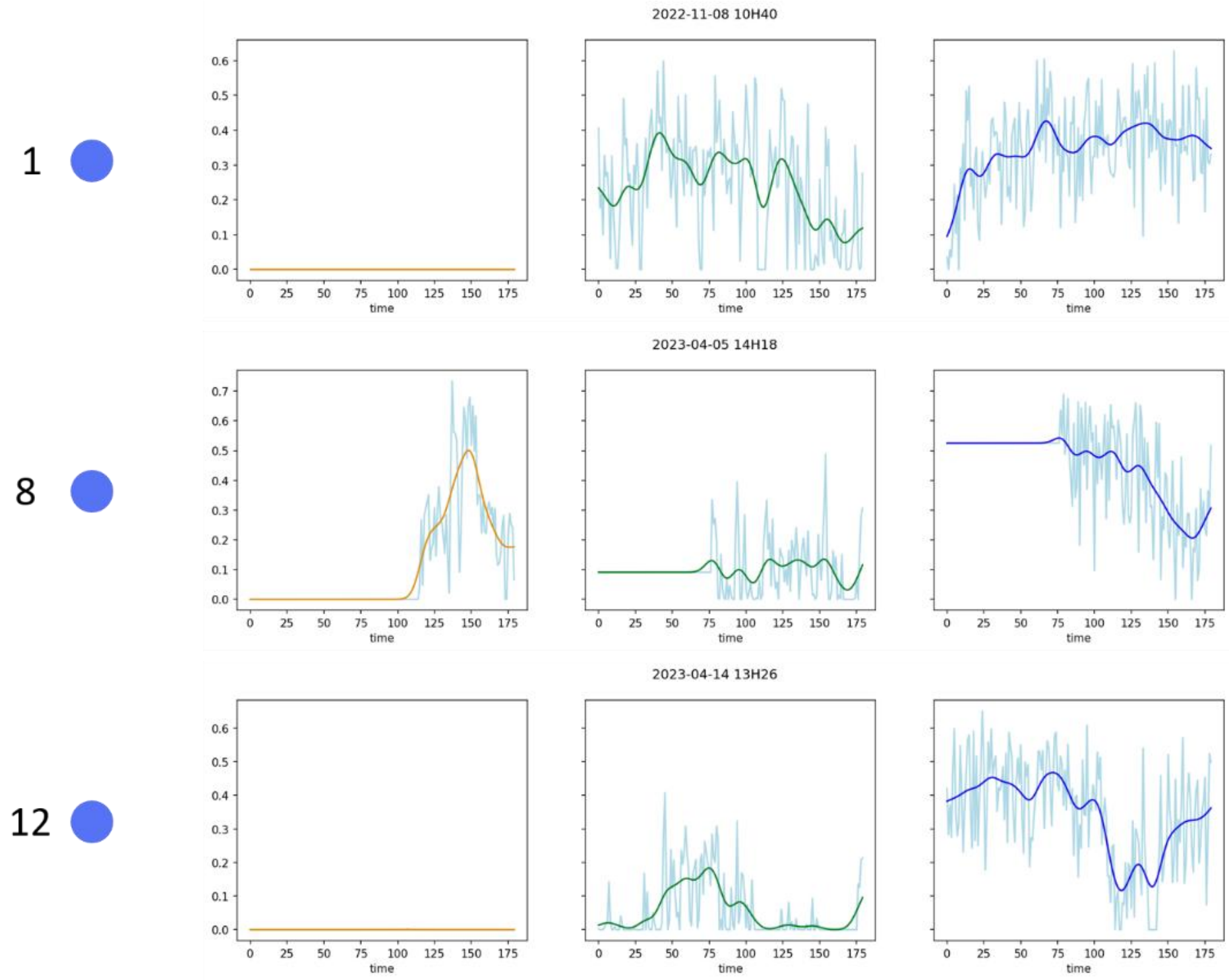

*Figure S10 Cluster 2 avoidance curves from field monitoring.*

**Cluster 3**

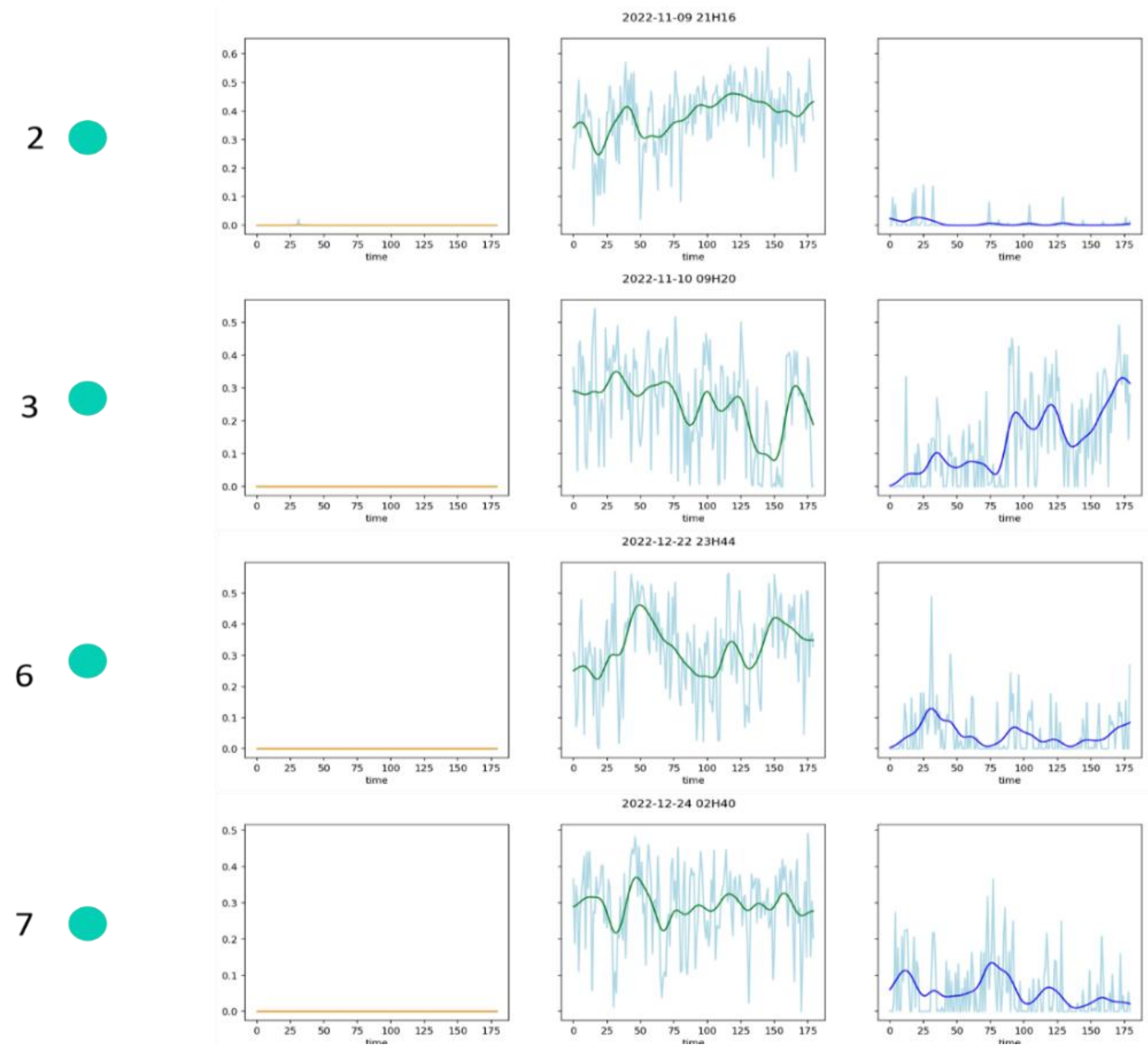

*Figure S11 Cluster 3 avoidance curves from field monitoring.*

**Cluster 4**

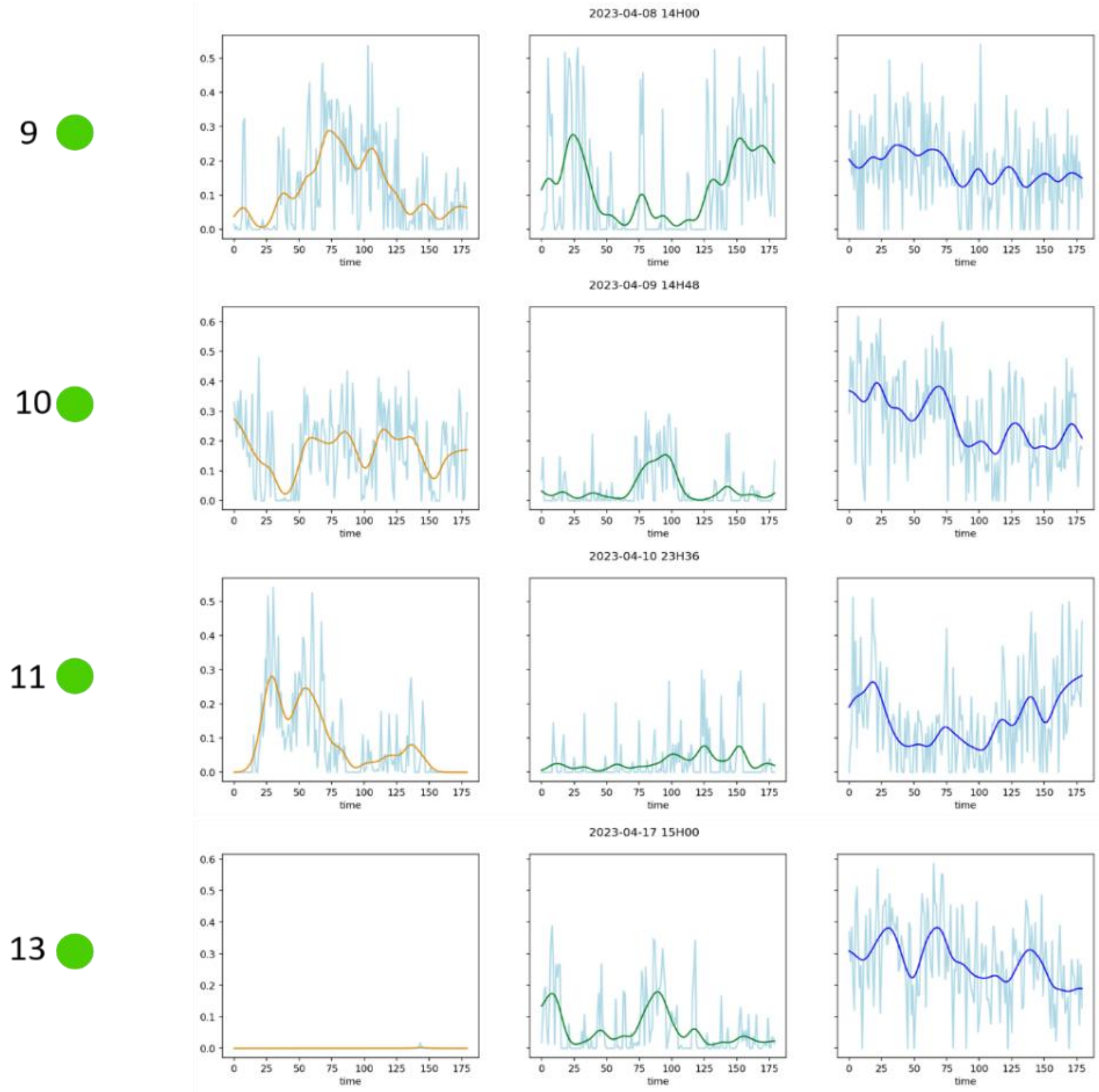

*Figure S12 Cluster 4 avoidance curves from field monitoring*

**Supplementary Table** – Details of the chemical screening of WWTP effluent samples shown on Figure 4.

| Estimated Concentration (µg/L) |
|---|
| < 0.05 |
| 0.05 à 0.1 |
| 0.1 à 0.5 |
| 0.5 à 1.0 |
| > 1.0 |

| Compound | Category | control | Sample 0 | Sample 1 | Sample 2 | Sample 3 | Sample 4 | Sample 5 | Sample 6 | Sample 7 | Sample 8 | Sample 9 | Sample 10 | Sample 11 | Sample 12 | Sample 13 |
|---|---|---|---|---|---|---|---|---|---|---|---|---|---|---|---|---|
| | Sampling date | 29/07/2022 11:30 | 16/08/2022 11:34 | 08/11/2022 10:40 | 09/11/2022 21:16 | 10/11/2022 09:20 | 10/11/2022 19:14 | 15/11/2022 12:20 | 22/12/2022 23:44 | 24/12/2022 02:40 | 05/04/2023 14:18 | 08/04/2023 14:00 | 09/04/2023 14:48 | 10/04/2023 23:36 | 14/04/2023 13:26 | 17/04/2023 15:00 |
| 10,11-Epoxycarbamazepine | pharmaceutical metabolite | | | 0.1 à 0.5 | | | | | | | 0.05 à 0.1 | 0.1 à 0.5 | 0.1 à 0.5 | Detected. Unquantified | Detected. Unquantified | 0.1 à 0.5 |
| 2-Aminopyridine | pharmaceutical | | | | | | | Detected. Unquantified | | | | | | | | |
| 5-methyl-1H-benzotriazole | other | | | | | | | | | | 0.5 à 1.0 | > 1.0 | > 1.0 | > 1.0 | > 1.0 | > 1.0 |
| Abacavir | pharmaceutical | | | | | | | < 0.05 | | | | | | | | |
| Acebutolol | pharmaceutical | < 0.05 | 0.05 à 0.1 | | | | | 0.05 à 0.1 | 0.05 à 0.1 | 0.05 à 0.1 | < 0.05 | 0.05 à 0.1 | < 0.05 | 0.05 à 0.1 | 0.05 à 0.1 | 0.05 à 0.1 |
| Acetamipride | pesticide | | | | | | | | | | | | | | Detected. Unquantified | Detected. Unquantified |
| Mefenamic acid | medicine | | Detected. Unquantified | | | | Detected. Unquantified | Detected. Unquantified | | | | | | | | |
| Amantadine | medicine | | | | 0.05 à 0.1 | | 0.1 à 0.5 | nan | 0.1 à 0.5 | 0.1 à 0.5 | < 0.05 | 0.05 à 0.1 | 0.05 à 0.1 | 0.05 à 0.1 | 0.05 à 0.1 | 0.05 à 0.1 |
| Amisulpride | pharmaceutical | | | | | | | | | | > 1.0 | > 1.0 | > 1.0 | > 1.0 | > 1.0 | > 1.0 |
| Amitriptyline | pharmaceutical | Detected. Unquantified | Detected. Unquantified | | | | | Detected. Unquantified | 0.1 à 0.5 | | < 0.05 | < 0.05 | < 0.05 | 0.05 à 0.1 | < 0.05 | 0.05 à 0.1 |
| Ammelide | pesticide metabolite | | | Detected. Unquantified | | | | | | | | | | | | |
| Atrazine | pesticide | < 0.05 | 0.05 à 0.1 | < 0.05 | < 0.05 | < 0.05 | < 0.05 | < 0.05 | < 0.05 | < 0.05 | < 0.05 | < 0.05 | < 0.05 | < 0.05 | < 0.05 | < 0.05 |
| Atrazine-2-hydroxy | pesticide metabolite | | < 0.05 | | | | | | | | | | | | | |
| Atrazine-desethyl | pesticide metabolite | < 0.05 | 0.1 à 0.5 | | < 0.05 | 0.05 à 0.1 | 0.05 à 0.1 | 0.05 à 0.1 | 0.05 à 0.1 | 0.05 à 0.1 | < 0.05 | < 0.05 | < 0.05 | < 0.05 | < 0.05 | < 0.05 |
| Azoxystrobin | pesticide | | < 0.05 | | | | | | | | | | | < 0.05 | | < 0.05 |
| Benomyl | pesticide | | | | | | | < 0.05 | | | | | | | | |
| Benzotriazole | other | | | | | | | | | | 0.5 à 1.0 | > 1.0 | > 1.0 | Detected. Unquantified | > 1.0 | 0.5 à 1.0 |
| Benzoylecgonine | pharmaceutical metabolite | | | | | | | | | | | | | Detected. Unquantified | | |
| Bezafibrate | pharmaceutical | | | | | | | | < 0.05 | 0.05 à 0.1 | | 0.05 à 0.1 | 0.05 à 0.1 | 0.05 à 0.1 | < 0.05 | 0.05 à 0.1 |
| Bisoprolol | pharmaceutical | Detected. Unquantified | Detected. Unquantified | | | | | Detected. Unquantified | Detected. Unquantified | Detected. Unquantified | 0.1 à 0.5 | 0.1 à 0.5 | 0.1 à 0.5 | 0.1 à 0.5 | 0.1 à 0.5 | 0.1 à 0.5 |
| Caffeine | pharmaceutical | | < 0.05 | 0.05 à 0.1 | < 0.05 | | | | | | | | | | | |
| Carbamazepine | pharmaceutical | 0.1 à 0.5 | > 1.0 | 0.5 à 1.0 | 0.1 à 0.5 | 0.5 à 1.0 | 0.5 à 1.0 | 0.5 à 1.0 | 0.5 à 1.0 | 0.5 à 1.0 | 0.1 à 0.5 | 0.5 à 1.0 | 0.5 à 1.0 | 0.5 à 1.0 | 0.5 à 1.0 | 0.5 à 1.0 |
| Carboxine | pesticide | | | | | | | < 0.05 | | | | | | | | |
| Ceftazidime | pharmaceutical | | | | | | | 0.1 à 0.5 | | | | | | | | |
| Cetirizine | pharmaceutical/pharmaceutical metab | 0.1 à 0.5 | > 1.0 | | | | | 0.5 à 1.0 | 0.1 à 0.5 | 0.1 à 0.5 | 0.1 à 0.5 | Detected. Unquantified | 0.5 à 1.0 | Detected. Unquantified | 0.5 à 1.0 | 0.5 à 1.0 |
| Chlortoluron | pesticide | | | | | | | < 0.05 | < 0.05 | < 0.05 | | < 0.05 | < 0.05 | < 0.05 | < 0.05 | |
| Cimetidine | pharmaceutical | | < 0.05 | | | | | | | | | | | | | |
| Citalopram | pharmaceutical | 0.1 à 0.5 | 0.1 à 0.5 | | | | | 0.1 à 0.5 | < 0.05 | 0.1 à 0.5 | 0.05 à 0.1 | 0.05 à 0.1 | 0.1 à 0.5 | 0.1 à 0.5 | 0.1 à 0.5 | 0.1 à 0.5 |
| Clindamycin | pharmaceutical | < 0.05 | < 0.05 | | | | | | | | | | | | | |
| Clomazone | pesticide | | | < 0.05 | | | | < 0.05 | | | | | | | | |
| Clopidogrel | pharmaceutical | < 0.05 | | | | < 0.05 | < 0.05 | < 0.05 | | < 0.05 | < 0.05 | < 0.05 | < 0.05 | < 0.05 | < 0.05 | < 0.05 |
| Clozapine | pharmaceutical | | | | | | | < 0.05 | | | | | | | | |
| Codeine | pharmaceutical | | | | | | | | | | Detected. Unquantified | Detected. Unquantified | Detected. Unquantified | Detected. Unquantified | Detected. Unquantified | Detected. Unquantified |
| Cyamemazine | pharmaceutical | | < 0.05 | | | | | < 0.05 | | | | | | | | |
| Cyclophosphamide | medicine | | | | | | | | < 0.05 | < 0.05 | | | | | < 0.05 | |
| Darunavir | pharmaceutical | Detected. Unquantified | | | | | Detected. Unquantified | | | | | | | | | |
| Desmethyldiazepam | pharmaceutical metabolite | Detected. Unquantified | | | Detected. Unquantified | Detected. Unquantified | Detected. Unquantified | | Detected. Unquantified | Detected. Unquantified | Detected. Unquantified | Detected. Unquantified | Detected. Unquantified | Detected. Unquantified | Detected. Unquantified | Detected. Unquantified |
| Desvenlafaxine | pharmaceutical/pharmaceutical metabolite | | | Detected. Unquantified | | | | | | | | | | | | |
| Dextromethorphan | pharmaceutical | | | | | | | | < 0.05 | < 0.05 | < 0.05 | < 0.05 | < 0.05 | | | |
| Dextrorphane | pharmaceutical/pharmaceutical metabolite | | < 0.05 | | | | < 0.05 | < 0.05 | | < 0.05 | < 0.05 | < 0.05 | < 0.05 | < 0.05 | < 0.05 | < 0.05 |
| Diazepam | pharmaceutical | | | | | | | | Detected. Unquantified | Detected. Unquantified | | | | | | |
| Diazinon | pesticide | | 0.5 à 1.0 | | | | | | | | | | | | | |
| Diclofenac | pharmaceutical | Detected. Unquantified | Detected. Unquantified | > 1.0 | > 1.0 | > 1.0 | Detected. Unquantified | > 1.0 | Detected. Unquantified | Detected. Unquantified | > 1.0 | > 1.0 | > 1.0 | > 1.0 | > 1.0 | > 1.0 |
| Diltiazem | pharmaceutical | | | | | | | | | | < 0.05 | < 0.05 | < 0.05 | < 0.05 | < 0.05 | < 0.05 |
| Dimetridazole | pharmaceutical | | | 0.05 à 0.1 | | < 0.05 | 0.05 à 0.1 | 0.05 à 0.1 | 0.05 à 0.1 | | | 0.05 à 0.1 | | | 0.05 à 0.1 | 0.1 à 0.5 |
| Diuron | pesticide | | < 0.05 | | | < 0.05 | | < 0.05 | | < 0.05 | < 0.05 | < 0.05 | | < 0.05 | < 0.05 | < 0.05 |
| Fenofibric acid | medicine | | | | | 0.1 à 0.5 | 0.05 à 0.1 | | | 0.5 à 1.0 | 0.1 à 0.5 | 0.1 à 0.5 | 0.1 à 0.5 | 0.5 à 1.0 | 0.1 à 0.5 | 0.1 à 0.5 |
| Fenuron | pesticide | | 0.1 à 0.5 | < 0.05 | | | | | | | | | | | | |
| Flubendazole | pharmaceutical | | | | | | | < 0.05 | < 0.05 | < 0.05 | < 0.05 | < 0.05 | < 0.05 | < 0.05 | < 0.05 | < 0.05 |
| Fluconazole | medicine | 0.05 à 0.1 | 0.1 à 0.5 | 0.1 à 0.5 | < 0.05 | | 0.1 à 0.5 | 0.1 à 0.5 | 0.1 à 0.5 | 0.1 à 0.5 | 0.05 à 0.1 | 0.1 à 0.5 | 0.1 à 0.5 | 0.1 à 0.5 | 0.1 à 0.5 | 0.1 à 0.5 |
| Flufenacet | pesticide | | | 0.05 à 0.1 | | | | | | | | | | | | |
| Fluopyram | pesticide | | < 0.05 | | | | | | | | | | | | | |
| Gabapentin | pharmaceutical | | | | | 0.1 à 0.5 | | Detected. Unquantified | 0.1 à 0.5 | 0.1 à 0.5 | 0.1 à 0.5 | 0.5 à 1.0 | | 0.5 à 1.0 | 0.5 à 1.0 | 0.5 à 1.0 |
| Imazalil | pesticide | | < 0.05 | | | | | < 0.05 | | | | | | | | |
| Imidacloprid | pesticide | | 0.05 à 0.1 | | | < 0.05 | | < 0.05 | < 0.05 | < 0.05 | | | | 0.05 à 0.1 | 0.05 à 0.1 | 0.05 à 0.1 |
| Irbesartan | pharmaceutical | | | | | | | | | | Detected. Unquantified | Detected. Unquantified | Detected. Unquantified | Detected. Unquantified | Detected. Unquantified | Detected. Unquantified |
| Isoproturon | pesticide | | | | | | | < 0.05 | | | | | | | | |
| Ketoprofen | pharmaceutical | | | | | < 0.05 | < 0.05 | < 0.05 | 0.05 à 0.1 | 0.1 à 0.5 | 0.05 à 0.1 | Detected. Unquantified | 0.05 à 0.1 | 0.1 à 0.5 | 0.1 à 0.5 | 0.1 à 0.5 |
| Levamisol | pharmaceutical | | | | | < 0.05 | < 0.05 | < 0.05 | | | | < 0.05 | < 0.05 | < 0.05 | < 0.05 | < 0.05 |
| …arbazepine (10,11-Dihydro-10-Hydroxycarbamazepine) | pharmaceutical metabolite | | | Detected. Unquantified | | | | | | | | | | | | |
| Lidocaine | pharmaceutical | | | Detected. Unquantified | Detected. Unquantified | Detected. Unquantified | Detected. Unquantified | | Detected. Unquantified | Detected. Unquantified | Detected. Unquantified | Detected. Unquantified | Detected. Unquantified | Detected. Unquantified | Detected. Unquantified | Detected. Unquantified |
| Lincomycin | pharmaceutical | | | | | | | | | | | | | | | |
| Lorazepam | pharmaceutical | | | Detected. Unquantified | | Detected. Unquantified | Detected. Unquantified | | Detected. Unquantified | Detected. Unquantified | | Detected. Unquantified | Detected. Unquantified | Detected. Unquantified | Detected. Unquantified | Detected. Unquantified |
| Losartan | pharmaceutical | | | | | | | | | < 0.05 | < 0.05 | < 0.05 | 0.05 à 0.1 | 0.05 à 0.1 | 0.05 à 0.1 | 0.05 à 0.1 |
| Loxapine | pharmaceutical | 0.05 à 0.1 | < 0.05 | | | | | < 0.05 | | | | | | | | |
| Melamine | pesticide metabolite | | Detected. Unquantified | Detected. Unquantified | Detected. Unquantified | | Detected. Unquantified | Detected. Unquantified | | | | Detected. Unquantified | Detected. Unquantified | Detected. Unquantified | Detected. Unquantified | Detected. Unquantified |
| Metazachlor | pesticide | | | < 0.05 | | | | | | | | | | | | |
| Metformin | pharmaceutical | | | | | | | | | | | | | 0.5 à 1.0 | 0.1 à 0.5 | 0.1 à 0.5 |
| Metoclopramide | pharmaceutical | | | | | < 0.05 | | < 0.05 | | < 0.05 | < 0.05 | < 0.05 | < 0.05 | < 0.05 | < 0.05 | < 0.05 |
| Metolachlor | pesticide | | < 0.05 | | | | | | | | | | | | | |
| Metoprolol | pharmaceutical | < 0.05 | 0.05 à 0.1 | | | | | 0.05 à 0.1 | 0.05 à 0.1 | 0.1 à 0.5 | < 0.05 | 0.05 à 0.1 | < 0.05 | < 0.05 | < 0.05 | < 0.05 |
| Metribuzin | pesticide | | | | | | | < 0.05 | | | | | | | | |
| Mianserin | pharmaceutical | < 0.05 | | | | | | | | | < 0.05 | < 0.05 | | | | |
| Myclobutanil | pesticide | < 0.05 | < 0.05 | | | | | | | | | | | 0.1 à 0.5 | 0.05 à 0.1 | < 0.05 |
| N-Ethylcyclohexylamine | other | | | | | | | | | | | > 1.0 | > 1.0 | Detected. Unquantified | Detected. Unquantified | Detected. Unquantified |
| Nortriptyline | pharmaceutical metabolite/pharmaceutical | | | | | | | < 0.05 | | | | | | | | |
| O-Desmethyltramadol | pharmaceutical metabolite | 0.1 à 0.5 | > 1.0 | 0.05 à 0.1 | 0.1 à 0.5 | > 1.0 | > 1.0 | > 1.0 | | 0.1 à 0.5 | 0.1 à 0.5 | 0.5 à 1.0 | > 1.0 | > 1.0 | > 1.0 | > 1.0 |
| Ofloxacin | pharmaceutical | 0.1 à 0.5 | 0.1 à 0.5 | 0.05 à 0.1 | 0.05 à 0.1 | | 0.05 à 0.1 | 0.05 à 0.1 | | | | | | | | |
| Omeprazole | pharmaceutical | | | | | | | 0.1 à 0.5 | 0.1 à 0.5 | 0.1 à 0.5 | 0.1 à 0.5 | 0.5 à 1.0 | 0.5 à 1.0 | 0.5 à 1.0 | 0.5 à 1.0 | 0.5 à 1.0 |
| Omeprazole sulfone | pharmaceutical metabolite | | | | | Detected. Unquantified | Detected. Unquantified | | Detected. Unquantified | Detected. Unquantified | | | | Detected. Unquantified | Detected. Unquantified | |
| Oxazepam | pharmaceutical/pharmaceutical metab | Detected. Unquantified | | Detected. Unquantified | Detected. Unquantified | Detected. Unquantified | Detected. Unquantified | | Detected. Unquantified | Detected. Unquantified | Detected. Unquantified | Detected. Unquantified | Detected. Unquantified | Detected. Unquantified | Detected. Unquantified | Detected. Unquantified |
| Oxycodone | narcotic | | Detected. Unquantified | | | | | Detected. Unquantified | | | | | | | | |
| Perindopril | pharmaceutical | < 0.05 | 0.05 à 0.1 | 0.05 à 0.1 | | 0.05 à 0.1 | | 0.05 à 0.1 | 0.05 à 0.1 | 0.05 à 0.1 | < 0.05 | 0.05 à 0.1 | 0.05 à 0.1 | 0.05 à 0.1 | 0.05 à 0.1 | 0.05 à 0.1 |
| Perindoprilate | pharmaceutical/pharmaceutical metabolite | | 0.1 à 0.5 | | | | | 0.05 à 0.1 | 0.1 à 0.5 | 0.1 à 0.5 | < 0.05 | 0.1 à 0.5 | 0.1 à 0.5 | 0.1 à 0.5 | 0.1 à 0.5 | 0.1 à 0.5 |
| Phenazone | pharmaceutical | < 0.05 | < 0.05 | < 0.05 | | < 0.05 | < 0.05 | | < 0.05 | < 0.05 | < 0.05 | Detected. Unquantified | < 0.05 | < 0.05 | < 0.05 | < 0.05 |
| Pheniramine | pharmaceutical | | | Detected. Unquantified | Detected. Unquantified | Detected. Unquantified | Detected. Unquantified | | Detected. Unquantified | Detected. Unquantified | Detected. Unquantified | Detected. Unquantified | Detected. Unquantified | Detected. Unquantified | Detected. Unquantified | Detected. Unquantified |
| Polymyxin B | pharmaceutical | | | Detected. Unquantified | | | | | | | | | | | | |
| Prazepam | pharmaceutical | | | Detected. Unquantified | Detected. Unquantified | Detected. Unquantified | Detected. Unquantified | | Detected. Unquantified | Detected. Unquantified | Detected. Unquantified | | Detected. Unquantified | Detected. Unquantified | Detected. Unquantified | Detected. Unquantified |
| Primidone | pharmaceutical | | 0.1 à 0.5 | 0.1 à 0.5 | | | 0.1 à 0.5 | 0.1 à 0.5 | | | < 0.05 | Detected. Unquantified | | 0.1 à 0.5 | 0.1 à 0.5 | 0.1 à 0.5 |
| Proguanil | medicine | < 0.05 | < 0.05 | | | | | < 0.05 | < 0.05 | | < 0.05 | | | | | |
| Propiconazole | pesticide | < 0.05 | < 0.05 | | | | | | | | < 0.05 | 0.05 à 0.1 | 0.05 à 0.1 | 0.05 à 0.1 | < 0.05 | < 0.05 |
| Propranolol | medicine | 0.05 à 0.1 | | | | | | 0.1 à 0.5 | 0.1 à 0.5 | 0.1 à 0.5 | 0.1 à 0.5 | 0.1 à 0.5 | 0.1 à 0.5 | 0.1 à 0.5 | 0.1 à 0.5 | 0.1 à 0.5 |
| Propyzamide | pesticide | | | | | | | | 0.05 à 0.1 | 0.1 à 0.5 | | | | | | |
| Prosulfocarb | pesticide | | | | | | | | | | Detected. Unquantified | Detected. Unquantified | | Detected. Unquantified | Detected. Unquantified | > 1.0 |
| Pseudoephedrine | pharmaceutical | | | Detected. Unquantified | | | | | | | | | | | | |
| Pyridostigmine | pharmaceutical | | | | | | | | | | | 0.1 à 0.5 | 0.1 à 0.5 | 0.1 à 0.5 | 0.1 à 0.5 | 0.1 à 0.5 |
| Ramipril | pharmaceutical | | | | | | | | | | | < 0.05 | < 0.05 | | < 0.05 | < 0.05 |
| Simazine | pesticide | < 0.05 | | < 0.05 | | | | < 0.05 | < 0.05 | < 0.05 | | | < 0.05 | < 0.05 | < 0.05 | < 0.05 |
| Sotalol | pharmaceutical | | 0.1 à 0.5 | 0.1 à 0.5 | 0.05 à 0.1 | 0.1 à 0.5 | 0.1 à 0.5 | 0.1 à 0.5 | 0.1 à 0.5 | 0.1 à 0.5 | 0.1 à 0.5 | 0.5 à 1.0 | 0.5 à 1.0 | 0.5 à 1.0 | 0.5 à 1.0 | 0.5 à 1.0 |
| Sulfamethoxazole | pharmaceutical | | 0.1 à 0.5 | 0.1 à 0.5 | | 0.1 à 0.5 | | 0.1 à 0.5 | 0.1 à 0.5 | 0.1 à 0.5 | 0.05 à 0.1 | 0.1 à 0.5 | 0.1 à 0.5 | 0.1 à 0.5 | 0.1 à 0.5 | 0.1 à 0.5 |
| Sulfapyridine | medicine | | 0.05 à 0.1 | | | | | 0.1 à 0.5 | | | | | | | | |
| Tebuconazole | pesticide | < 0.05 | | < 0.05 | | | | | | | | | | | | |
| Tebuthiuron | pesticide | | | | | | | < 0.05 | | | | | | | | |
| Temazepam | pharmaceutical | Detected. Unquantified | | | | Detected. Unquantified | | | Detected. Unquantified | Detected. Unquantified | Detected. Unquantified | Detected. Unquantified | Detected. Unquantified | Detected. Unquantified | Detected. Unquantified | Detected. Unquantified |
| Terbutryne | pesticide | < 0.05 | < 0.05 | < 0.05 | | < 0.05 | < 0.05 | < 0.05 | | | | Detected. Unquantified | < 0.05 | < 0.05 | < 0.05 | < 0.05 |
| Tertbuthylazine-2-hydroxy | Pesticide metabolite | | | | | | | | | | < 0.05 | < 0.05 | < 0.05 | < 0.05 | < 0.05 | < 0.05 |
| Theophylline | pharmaceutical | | | | | | | | | | 0.05 à 0.1 | 0.1 à 0.5 | | 0.1 à 0.5 | 0.05 à 0.1 | 0.1 à 0.5 |
| Thiabendazole | other | < 0.05 | < 0.05 | < 0.05 | < 0.05 | < 0.05 | < 0.05 | < 0.05 | < 0.05 | < 0.05 | < 0.05 | < 0.05 | < 0.05 | < 0.05 | < 0.05 | < 0.05 |
| Timolol | pharmaceutical | | < 0.05 | | | | | < 0.05 | | | | | | | | |
| Tramadol | pharmaceutical | Detected. Unquantified | > 1.0 | Detected. Unquantified | 0.5 à 1.0 | > 1.0 | > 1.0 | > 1.0 | > 1.0 | > 1.0 | 0.5 à 1.0 | > 1.0 | > 1.0 | > 1.0 | > 1.0 | > 1.0 |
| Trimethoprim | pharmaceutical | | < 0.05 | | | | < 0.05 | < 0.05 | < 0.05 | < 0.05 | < 0.05 | 0.05 à 0.1 | 0.05 à 0.1 | < 0.05 | 0.05 à 0.1 | < 0.05 |
| Valsartan | medication | | Detected. Unquantified | | | | 0.5 à 1.0 | Detected. Unquantified | | > 1.0 | > 1.0 | > 1.0 | > 1.0 | > 1.0 | > 1.0 | > 1.0 |
| Venlafaxine | medication | 0.1 à 0.5 | | | | | | 0.1 à 0.5 | 0.1 à 0.5 | 0.1 à 0.5 | 0.1 à 0.5 | 0.5 à 1.0 | 0.5 à 1.0 | 0.5 à 1.0 | 0.5 à 1.0 | 0.5 à 1.0 |

*Summary for highest concentrations*

| Sample 1 | | Sample2 | | Sample3 | | Sample4 | | Sample5 | | Sample6 | | Sample7 | | Sample8 | | Sample9 | | Sample10 | | Sample11 | | Sample12 | | Sample13 | |
|---|---|---|---|---|---|---|---|---|---|---|---|---|---|---|---|---|---|---|---|---|---|---|---|---|---|
| **Chemical** | **Concentration** | **Chemical** | **Concentration** | **Chemical** | **Concentration** | **Chemical** | **Concentration** | **Chemical** | **Concentration** | **Chemical** | **Concentration** | **Chemical** | **Concentration** | **Chemical** | **Concentration** | **Chemical** | **Concentration** | **Chemical** | **Concentration** | **Chemical** | **Concentration** | **Chemical** | **Concentration** | **Chemical** | **Concentration** |
| carbamazepine | 0.75 | tramadol | 0.75 | tramadol | 1.5 | tramadol | 1.5 | tramadol | 1.5 | tramadol | 1.5 | tramadol | 1.5 | tramadol | 0.75 | gabapentine | 0.75 | sotalol | 0.75 | metformine | 0.75 | gabapentine | 0.75 | gabapentine | 0.75 |
| diclofenac | 1.5 | diclofenac | 1.5 | carbamazepine | 0.75 | valsartan | 0.75 | carbamazepine | 0.75 | carbamazepine | 0.75 | valsartan | 1.5 | valsartan | 1.5 | sotalol | 0.75 | tramadol | 1.5 | gabapentine | 0.75 | sotalol | 0.75 | sotalol | 0.75 |
| | | | | o-desmethyltran | 1.5 | o-desmethyltran | 1.5 | diclofenac | 1.5 | | | carbamazepine | 0.75 | amisulpride | 1.5 | tramadol | 1.5 | valsartan | 1.5 | sotalol | 0.75 | tramadol | 1.5 | tramadol | 1.5 |
| | | | | diclofenac | 1.5 | | | cetirizine | 0.75 | | | fenofibrique acic | 0.75 | diclofenac | 1.5 | valsartan | 1.5 | venlafaxine | 0.75 | tramadol | 1.5 | valsartan | 1.5 | valsartan | 1.5 |
| | | | | | | | | o-desmethyltran | 1.5 | | | | | benzotriazole | 0.75 | venlafaxine | 0.75 | omeprazole | 0.75 | valsartan | 1.5 | venlafaxine | 0.75 | venlafaxine | 0.75 |
| | | | | | | | | | | | | | | 5-methyl-1h-ber | 0.75 | omeprazole | 0.75 | carbamazepine | 0.75 | venlafaxine | 0.75 | omeprazole | 0.75 | omeprazole | 0.75 |
| | | | | | | | | | | | | | | | | carbamazepine | 0.75 | amisulpride | 1.5 | omeprazole | 0.75 | carbamazepine | 0.75 | carbamazepine | 0.75 |
| | | | | | | | | | | | | | | | | amisulpride | 1.5 | diclofenac | 1.5 | carbamazepine | 0.75 | amisulpride | 1.5 | amisulpride | 1.5 |
| | | | | | | | | | | | | | | | | diclofenac | 1.5 | cetirizine | 0.75 | amisulpride | 1.5 | diclofenac | 1.5 | diclofenac | 1.5 |
| | | | | | | | | | | | | | | | | o-desmethyltran | 0.75 | o-desmethyltran | 1.5 | diclofenac | 1.5 | cetirizine | 0.75 | cetirizine | 0.75 |
| | | | | | | | | | | | | | | | | benzotriazole | 1.5 | benzotriazole | 1.5 | fenofibrique acic | 0.75 | o-desmethyltran | 1.5 | o-desmethyltran | 1.5 |
| | | | | | | | | | | | | | | | | n-ethylcyclohex | 1.5 | n-ethylcyclohex | 1.5 | o-desmethyltran | 1.5 | benzotriazole | 1.5 | prosulfocarbe | 1.5 |
| | | | | | | | | | | | | | | | | 5-methyl-1h-ber | 1.5 | 5-methyl-1h-ber | 1.5 | 5-methyl-1h-ber | 1.5 | 5-methyl-1h-ber | 1.5 | benzotriazole | 0.75 |
| | | | | | | | | | | | | | | | | | | | | | | | | 5-methyl-1h-ber | 1.5 |